\documentclass[prl,twocolumn,showpacs,amsmath,amssymb,superscriptaddress,floatfix]{revtex4}

\usepackage[english]{babel}
\usepackage{amsfonts}
\usepackage{graphicx}
\usepackage{times}
\usepackage[normalem]{ulem} 

\usepackage{color}
\definecolor{nred} {RGB}{224,0,0}
\definecolor{nblue} {RGB}{28,130,185}
\definecolor{dgreen} {RGB}{78,138,21}

\newcommand{\mat}[1]{\mathrm{#1}}

\begin{document}

\title{Fingerprints of preformed pairs  in two-electron angle-resolved photoemission spectroscopy}

\author{Janez \surname{Bon\v ca}}
\affiliation{Faculty of Mathematics and Physics, University of Ljubljana, 1000
Ljubljana, Slovenia}
\affiliation{J. Stefan Institute, 1000 Ljubljana, Slovenia}

\author{Andrea \surname{Damascelli}}
\affiliation{Department of Physics and Astronomy, University of British Columbia, Vancouver, BC, Canada, V6T 1Z1}
\affiliation{Quantum Matter Institute, University of British Columbia, Vancouver, BC, Canada, V6T 1Z4}

\author{Mona \surname{Berciu}}
\affiliation{Department of Physics and Astronomy, University of British Columbia, Vancouver, BC, Canada, V6T 1Z1}
\affiliation{Quantum Matter Institute, University of British Columbia, Vancouver, BC, Canada, V6T 1Z4}

\date{\today}
\begin{abstract}
We use variational exact diagonalization (VED) to
calculate the two-electron removal spectral weight for the Hubbard-Holstein model, starting from the ground-state with two electrons on a one-dimensional chain. We argue that this spectral weight provides a valuable proxy for the intensity of 2eARPES processes. Our results show that when contrasted to the presumably larger signal due to two electrons ejected from two different pairs, the presumably weaker signal due to two electrons ejected from the same pair  (i) is segregated in energy, appearing at a lower binding energy, and (ii) has a very characteristic momentum dependence, with a different symmetry than that of the signal corresponding to two electrons emitted from two different pairs. We verify that these fingerprints appear for pairs with different symmetries, and prove that they arise as a direct consequence of momentum and energy conservation, therefore they are generic for any model with electron-boson coupling that can lead to formation of electron pairs. Experimental observation of these fingerprints will confirm the existence of pairs. Moreover, the momentum dependence map allows one to distinguish whether the pairs are coherent (superconducting) or not. Finally, we argue that these considerations generalize to finite but low electron concentrations, finite temperatures and higher dimensions.
\end{abstract}  

\maketitle

\setcounter{figure}{0}

{\em Introduction:} Angle-resolved photoemission spectroscopy (ARPES) is a well-established and  extremely successful experimental technique that provides direct access to the quasiparticle band structure and Fermi surfaces of a variety of interesting materials \cite{Damascelli2003}. Particularly useful for strongly-correlated electron systems is the fact that ARPES also enables access to  the quasiparticle self-energy. This quantity can be compared against self-energies corresponding to various correlated model Hamiltonians, allowing theorists to gain insights into the most relevant physics of the system of interest.

While access to these single-particle properties is extremely valuable, it is not sufficient to fully characterize the state of a  correlated system. For example, direct signatures of pair formation, {\em e.g.} in the context of superconductivity, is seen with ARPES  if the superconducting gap opened in the single-particle density of states is within the energy resolution of the ARPES system. However, generically ARPES cannot  identify whether the gap is due to superconductivity or some other order;  additional measurements are needed to settle that \cite{Damascelli2004}.

This becomes problematic  when trying to identify new  correlated states that are not well understood theoretically. Of primary interest are liquids of pre-formed pairs, hypothesized to appear in systems where boson exchange  favours pairing of the electrons, however these pairs are not coherent (the system is not superconducting) because the electron density is too low or the temperature is too high. We recently argued that signatures of such a state could be seen in ARPES \cite{Klemen2025}, however it  is recognized that one of the most direct diagnostics for pairing correlations is provided by two electron coincidence ARPES (2eARPES); other possible correlation-focussed spectroscopies are reviewed in Ref. \cite{Su2025}. Indeed, analysis of the probability of coincidence detection of a pair of ejected electrons upon absorption of one photon has  allowed the observation of the  exchange-correlation hole surrounding an electron in a correlated (but unpaired) metal \cite{Kirschner_2007}, while recent theoretical work  argued that 2eARPES can be used to identify the pairing state in BCS superconductors whose Cooper pairs have a nonzero center-of-mass momentum \cite{T2}, and to extract dynamic electron
interactions in other ordered phases, such as a charge density wave \cite{Kemper2025}. Exact calculation of two-photons in, two-electrons out 2eARPES intensities has revealed clear signatures of pre-formed pair formation for  an 8-site cluster attractive Hubbard model at half-filling \cite{Devereaux_2023}.

In this Letter, we  use the formalism developed in Refs. \cite{Eckstein2019,Su2020,Devereaux_2023} to calculate 2eARPES intensities. First, we argue that when compared to the presumably larger signal due to two electrons ejected from two different pairs,  the signal due to two electrons ejected from the same pair is always (i) segregated in energy, and (ii) has a characteristic momentum-dependence. These fingerprints are a direct consequence of momentum and energy conservation, therefore they must appear in any model hosting pre-formed pairs.  Their observation  in 2eARPES will directly confirm the presence of pairs, allow for an estimate of their real-space radius, and also indicate whether the pairs are coherent or not. We then verify this  by calculating numerically the 2eARPES intensity of a pre-formed pair (a bipolaron) in the one-dimensional (1D) Hubbard-Holstein model. Varying the electron-phonon coupling strength and the on-site Coulomb repulsion, we can contrast the results for ground-states (GS) consisting of two unbound polarons (strong repulsion, weak el-ph coupling) from those for GS with a bipolaron (weak repulsion, strong el-ph coupling). We confirm that the two fingerprints are observed both for $s$- and $p$-wave pairs.


{\em Model and Method:} We study  an $N$-site 1D chain  with electrons that experience Hubbard repulsion and are  also Holstein-coupled to dispersive optical phonons, as described by the Hamiltonian:
\begin{eqnarray} \label{hol}
\vspace*{-0.0cm}
H &=& -t_\mathrm{el}\sum_{j,\sigma}(c^\dagger_{j,\sigma} c_{j+1,\sigma} +\mathrm{H.c.}) + {g} \sum_{j} \hat n_{j} (a_{j}^\dagger + a_{j})  \nonumber \\
 &+&  t_\mathrm{ph}\sum_{j}(a^\dagger_{j} a_{j+1} +\mathrm{H.c.})
+ \omega_0\sum_{j} a_{j}^\dagger  a_{j} \nonumber\\
&+& U\sum_j n_{j\uparrow}n_{j\downarrow}, \label{ham}
\end{eqnarray}
where $c^\dagger_{j\sigma}$ and $a^\dagger_{j}$ are electron and phonon creation operators at site $j$, respectively,  $\hat n _{j} = \sum_\sigma c^\dagger_{j\sigma}c_{j\sigma}$  is the density operator and $U$ is the Hubbard repulsion. The free-electron dispersion $\epsilon_k = - 2 t_{el} \cos(k)$ is controlled by the nearest-neighbor hopping amplitude taken as the energy unit $t_\mat{el}\equiv 1$, while the optical phonon energy $\Omega_q=\omega_0 + 2t_\mathrm{ph}\cos(q)$ is such that $\omega_0>  |2t_{ph}|$. We set the lattice constant $a=1$, also $\hbar =1$. The electron-phonon  coupling strength is hereafter characterized by 
 the  dimensionless  $\lambda =g^2/[2t_\mat{el}\sqrt{\omega_0^2-4t_\mathrm{ph}^2}]$  \cite{berciu2013}.  

We study this Hamiltonian for $N_e=2$ electrons on a finite chain with periodic boundary conditions employing variational exact diagonalization (VED) \cite{bonca1,bonca2000prl,bonca2002,bonca2021,Klemen2024,KlemenPRB2025}. We  typically use $N_h\sim 18$ iterations to generate the variational space by repeated application of $H$; we verified that the results are converged and correspond to the thermodynamic limit $N \rightarrow \infty$. 
 
{\em 2eARPES Intensity:} We calculate the spectral weight $A_2(\omega,k_1,k_2) = {1\over \pi} \mbox{Im} {\cal G}_{k_1k_2}(\omega)$ of the  two-particle propagator:
\begin{equation}
    {\cal G}_{k_1k_2}(\omega) = \sum_{n} \frac{|\langle \psi^{(n,N_e-2)}_{ -k_1-k_2}| c_{k_1\sigma_1 }c_{k_2\sigma_2}| \psi^{(GS,N_e)}_0\rangle|^2}{\omega - i \eta +E_{-k_1-k_2}^{(n,N_e-2)}-E_0^{(GS,N_e)}}
    \label{G}
\end{equation}
Here, ${\cal H}  = |\psi_k^{(n,N_e)}\rangle =E_k^{(n,N_e)}|\psi_k^{(n,N_e)}\rangle$ are the eigenstates and eigenenergies with $N_e$ particles and total momentum $k$; $n$ labels other quantum numbers. The ground-state GS corresponds to $n=0$ and has total momentum 0. 
For results in the singlet sector we set $\sigma_1=-\sigma_2$ and verify that double occupancy has a non-vanishing probability. For the triplet sector, we set $\sigma_1=\sigma_2$. In this case double occupancy is forbidden by the Pauli principle, and the value of $U$ becomes irrelevant.

\begin{widetext}
This two-particle propagator can be linked directly to 
the quantity defined in Eq. (23) of Ref. \onlinecite{Devereaux_2023}:
\begin{equation}
    D^{(0)}_{{\bf k_1 k_2}}(\omega) \!= \!\int_{-\infty}^{\infty}\!\!\! dt e^{-i\omega t}\!\!\int_{-\infty}^{\infty}\!\!\!\! d\tau \langle c^\dagger_{k_1\sigma_1}(t) c^\dagger_{k_2\sigma_2}(t+\tau) c_{k_2\sigma_2}(\tau) c_{k_1\sigma_1}(0)
\rangle
\label{Dev}
\end{equation}
\end{widetext}
Up to a factor containing photoexcitation matrix elements,  which is  ignored for simplicity, $D^{(0)}_{{\bf k_1 k_2}}(\omega)$ was shown by Devereaux {\em et al.} \cite{Devereaux_2023} to represent the 2eARPES coincidence detection rate integrated over all possible differences $\Delta \omega =\omega_1 -\omega_2$ at a fixed total $\omega = \omega_1+\omega_2$, where $\omega_i= \omega_{ph} - {\hbar^2 {\bf k}_i^2\over 2m}-W$ is the energy imparted to  the system upon absorption of photon $i$ and emission of the photo-electron with momentum ${\bf k_i}$. We denote ${\bf k}_i= k_i{\bf x} + {\bf k}_{i,\perp}$, {\em i.e.} the conserved (quasi)momentum along the chain is called $k_i$ (note that $k_i \ne |{\bf k}_i|$); this is why the electron  operators associated with the  states involved in the photoemission processes have (quasi)momenta $k_1,k_2$. Finally, $W$ is the work function and the expectation value $\langle \dots \rangle$ is over the GS with $N_e$ particles. Physically, $\tau$ is the time difference between the emission of the first and the second electron of the detected 2e ARPES pair \cite{Devereaux_2023}. 

For reasons that will be justified {\em a posteriori}, 
in the following we focus on the simpler case $\tau \approx 0$, when Eq. (\ref{Dev}) simplifies straightforwardly to $D^{(0)}_{{\bf k_1 k_2}}(\omega) \propto A(k_1,k_2,\omega)$ (the dependence on the transverse momenta of the photo-electrons is through the total energy $\omega$ imparted to the system).

Next, we analyze Eq. (\ref{G}) to infer the generic properties of the expected 2eARPES intensity  for a ground-state of preformed pairs that are  very weakly interacting with one another.  In Ref. \onlinecite{Klemen2025} we showed that such a liquid of $s$-like singlet bipolarons is the GS of the 1D Hubbard-Holstein model at low carrier concentrations, if the coupling $\lambda$ is sufficiently large and the repulsion $U$ is sufficiently small. Here, the pre-formed pairs are bipolarons of energy $E_{BP}(K=0)= 2E_P(0)-\Delta$ (the bipolaron GS momentum is $K=0$), where $\Delta$ is the binding energy and $E_P(k)$ is the energy of a single polaron with momentum $k$ (the single polaron GS is also at $k=0$). Interestingly, this model also allows for the binding of $p$-like, triplet bipolarons if  $t_{ph}<0$. As shown in Ref. \cite{Klemen2024}, dispersive phonons mediate an effective nearest-neighbor interaction $\Delta E = 2 t_{ph} g^2/\omega_0^2$ which becomes attractive if $t_{ph}<0$ and suffices to bind triplet pairs when $\Delta E$ is sufficiently negative, see the Supplementary Material for more details \cite{SM}.

Equation (\ref{G}) shows that 2eARPES spectral weight is expected at the energy $\omega = E_0^{(GS,N_e)} - E_{-k_1-k_2}^{(\alpha,N_e-2)}$. First we consider the case where the two electrons originate in the same bipolaron. Then $E_0^{(GS,N_e)}\approx E_{BP}(0) + E_0^{(GS,N_e-2)}$ due to the weak bipolaron interactions. The lowest binding energy feature corresponds to $E_{-k_1-k_2}^{(\alpha,N_e-2)} = E_0^{(GS,N_e-2)}$, {\em i.e.} the bipolaron is removed and no phonons are left behind. This process is only possible if $k_1+k_2=0$ and has energy $\omega = E_{BP}(0)=2\mu$ \cite{Klemen2025}, confirming that this bipolaron liquid has no gap to pair excitations, $\omega - 2\mu \le 0$. The next feature appears at $\omega -2\mu = - \Omega_{-k_1-k_2}$ if the 2eARPES process removes the bipolaron but leaves behind one phonon carrying the total momentum imparted to the system; this is followed by a two-phonon continuum when two phonons that share the total momentum $-k_1-k_2$ are left behind, etc. 

This process (both electrons ejected from the same pair) competes with the much more probable process of the two electrons being ejected from two different pairs: the latter's probability scales like $N_p(N_p-1)$ where $N_p=N_e/2$ is the number of pairs, while the former's scales like $N_p$. This is why we assume throughout  that the latter's contribution to the 2eARPES intensity  is  presumably larger than the former's. 

Consider  having  two electrons ejected from two pairs, starting from $E_0^{(GS,N_e)}\approx 2 E_{BP}(0) + E_0^{(GS,N_e-4)}$. The lowest energy eigenstate upon removing an electron from a bipolaron corresponds to formation of a polaron, followed by higher energy replicas when phonons are also left behind. It follows that  $E_{-k_1-k_2}^{(\alpha,N_e-2)} \ge E_P(-k_1)+ E_P(-k_2) +E_0^{(GS,N_e-4)} $ (we ignore weak interactions between the two polarons and the remaining bipolarons, which is reasonable at very low densities). If the two  polarons are far apart, they cannot instantaneously bind into a new bipolaron, thus the  lowest binding energy feature  is at $\omega \le 2 E_{BP}(0) - E_P(-k_1)- E_P(-k_2) \le 2E_{BP}(0) - 2E_P(0)= 2\mu  - \Delta $, {\em i.e.} {\it at  energy $\Delta$ below the lowest binding energy feature for two electrons ejected from the same pair}.  This energy separation allows the presumably weaker 2eARPES signal originated from  the same pair to be distinguished from the presumably stronger  signal from two electrons emitted from two pairs. 

This prediction is reminiscent of the results presented in Ref. \onlinecite{T2} for a BCS superconductor, which also shows the intensity due to both electrons coming from the same Cooper pair to be energetically separated from that for electrons emitted from different Cooper pairs. Here,  we confirm that this energy separation between the two signals also appears for a liquid of incoherent pre-formed pairs, and that it is an exact result, not a mean-field/BCS approximation.

{\em Results:} We now confirm these general expectations using VED for $N_e=2$, both for singlet $s$- and triplet $p$-symmetry pairs, by calculating  $A_2(\omega,k_1,k_2)$ from Eq. (\ref{G}). We find that  $50$ Lanczos steps typically sufficed to obtain  accurately the bipolaron GS wavefunction and energy.   We truncated the sum over the $N_e-2=0$ eigenstates  in Eq. (\ref{G}) at $n \le n_{max}=200$ and employed the Gram-Schmid reorthogonalization procedure to ensure their orthogonality. An important test  for the accuracy of the numerical results is the sum rule $\sum_{k_1,k_2}\int_{-\infty}^{+\infty}d\omega A_2(\omega,k_1,k_2)=1$ (in the singlet sector) and $2$ (in the triplet sector). For all results shown here, this sum rule is satisfied to  four significant digits.

 \begin{figure}[t]
\includegraphics[width=0.5\columnwidth]{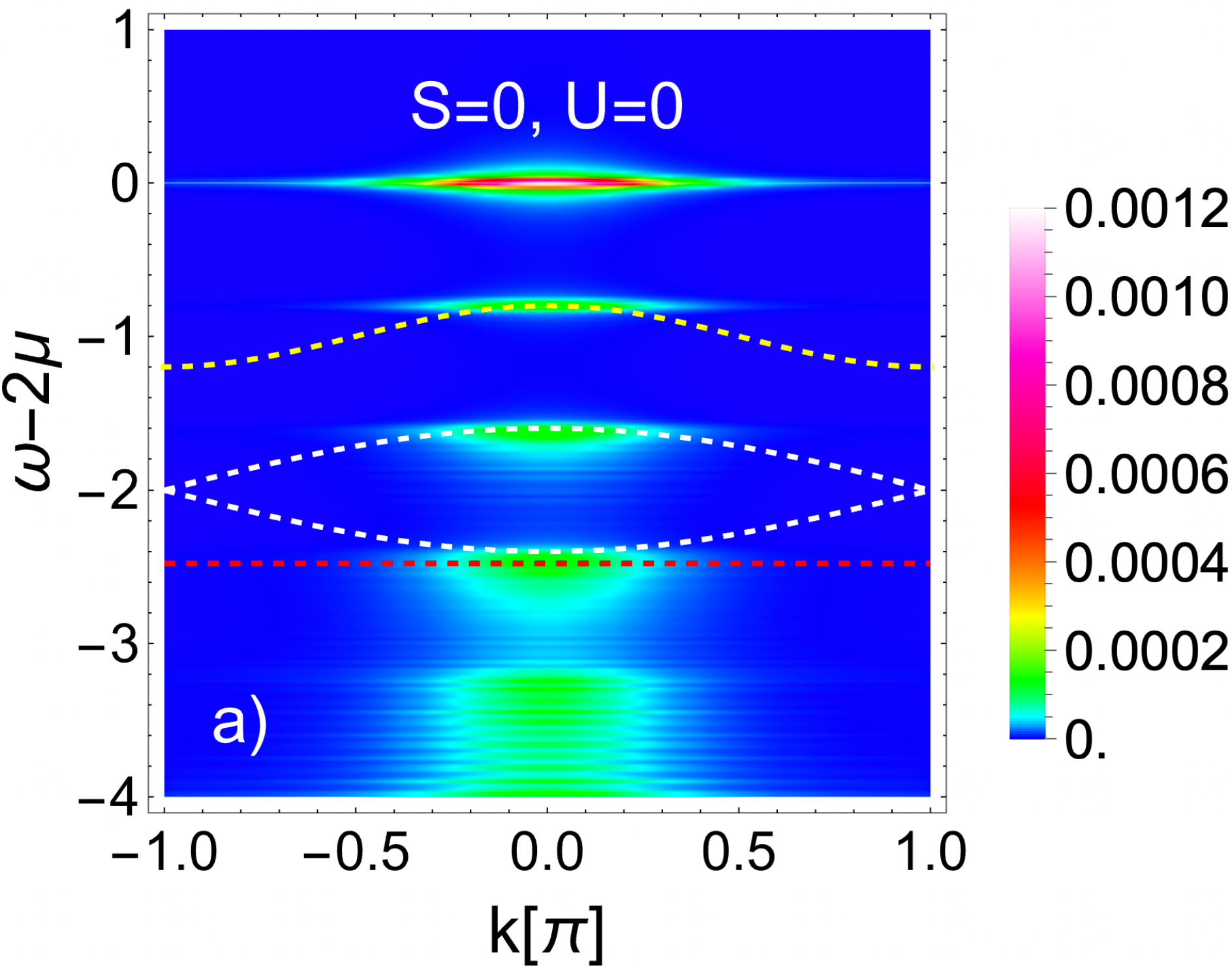}\includegraphics[width=0.5\columnwidth]{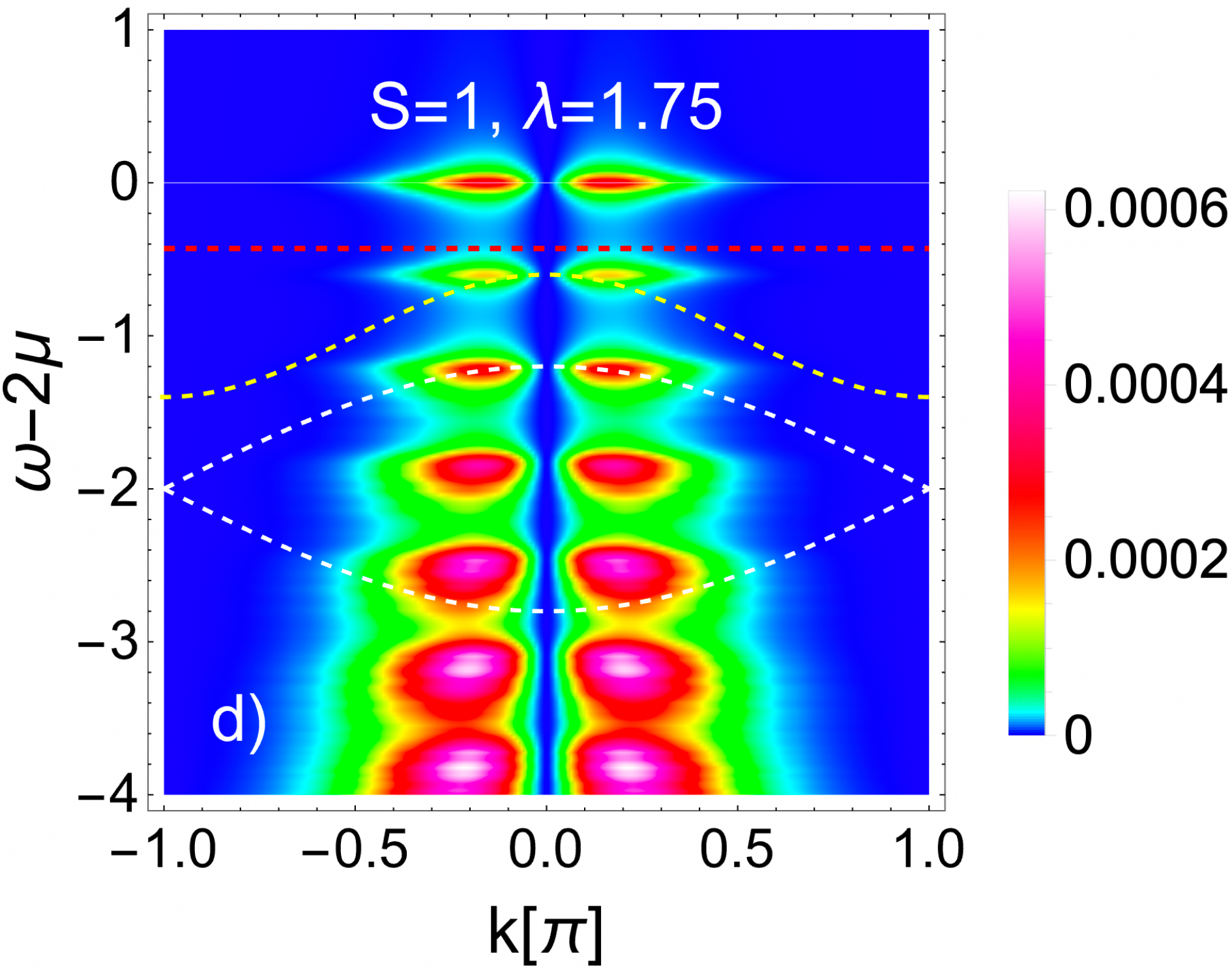}
\includegraphics[width=0.5\columnwidth]{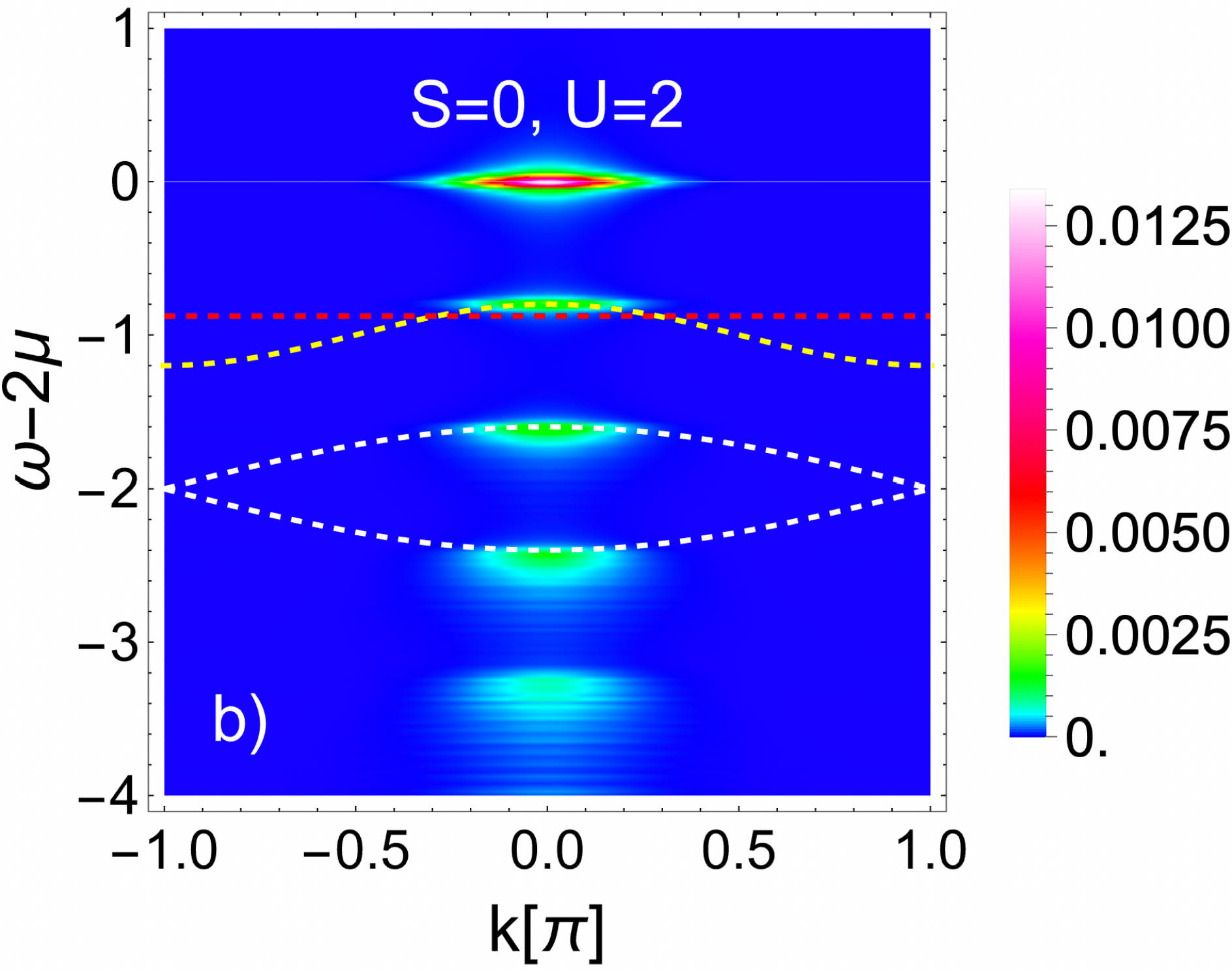}\includegraphics[width=0.5\columnwidth]{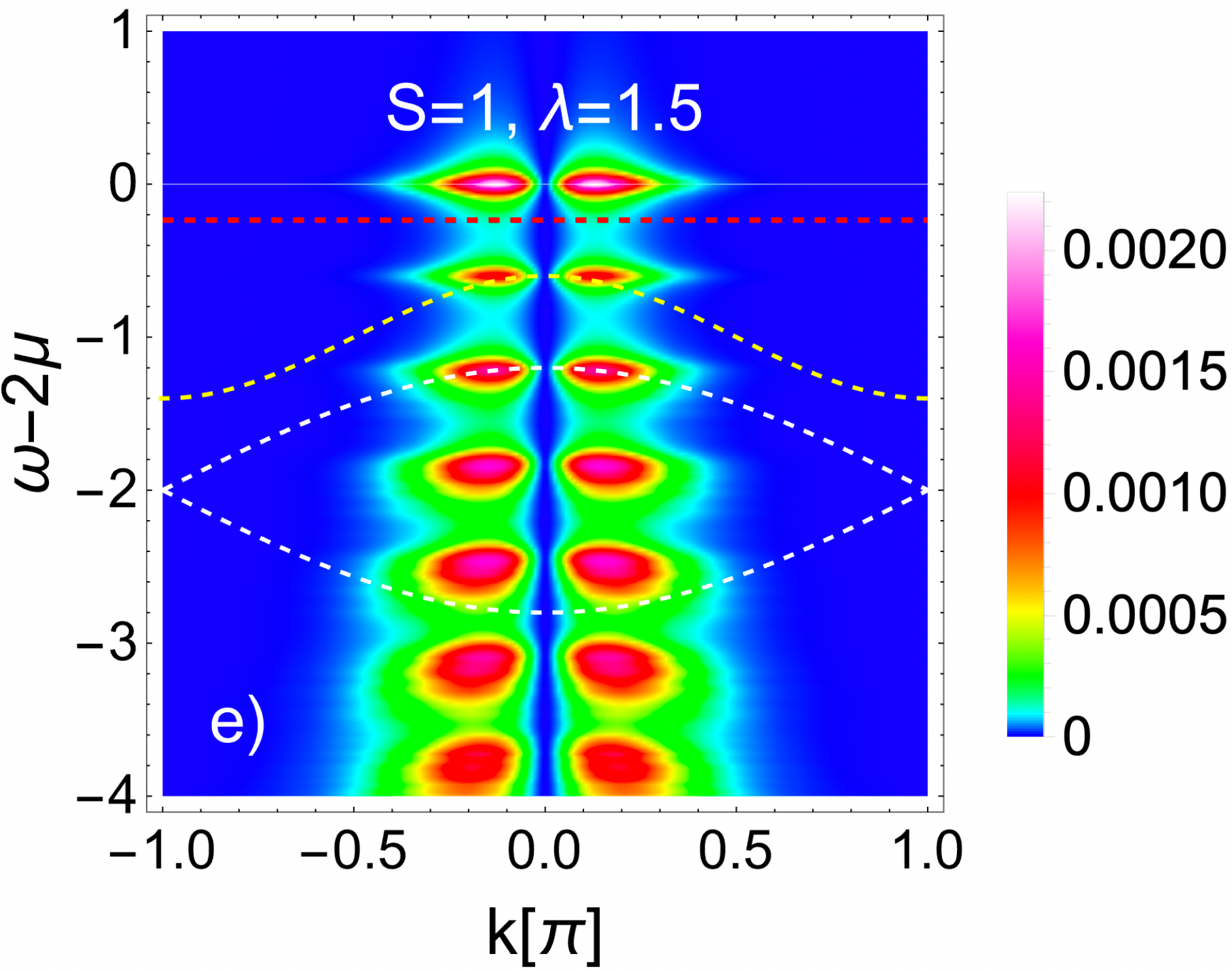}
\includegraphics[width=0.5\columnwidth]{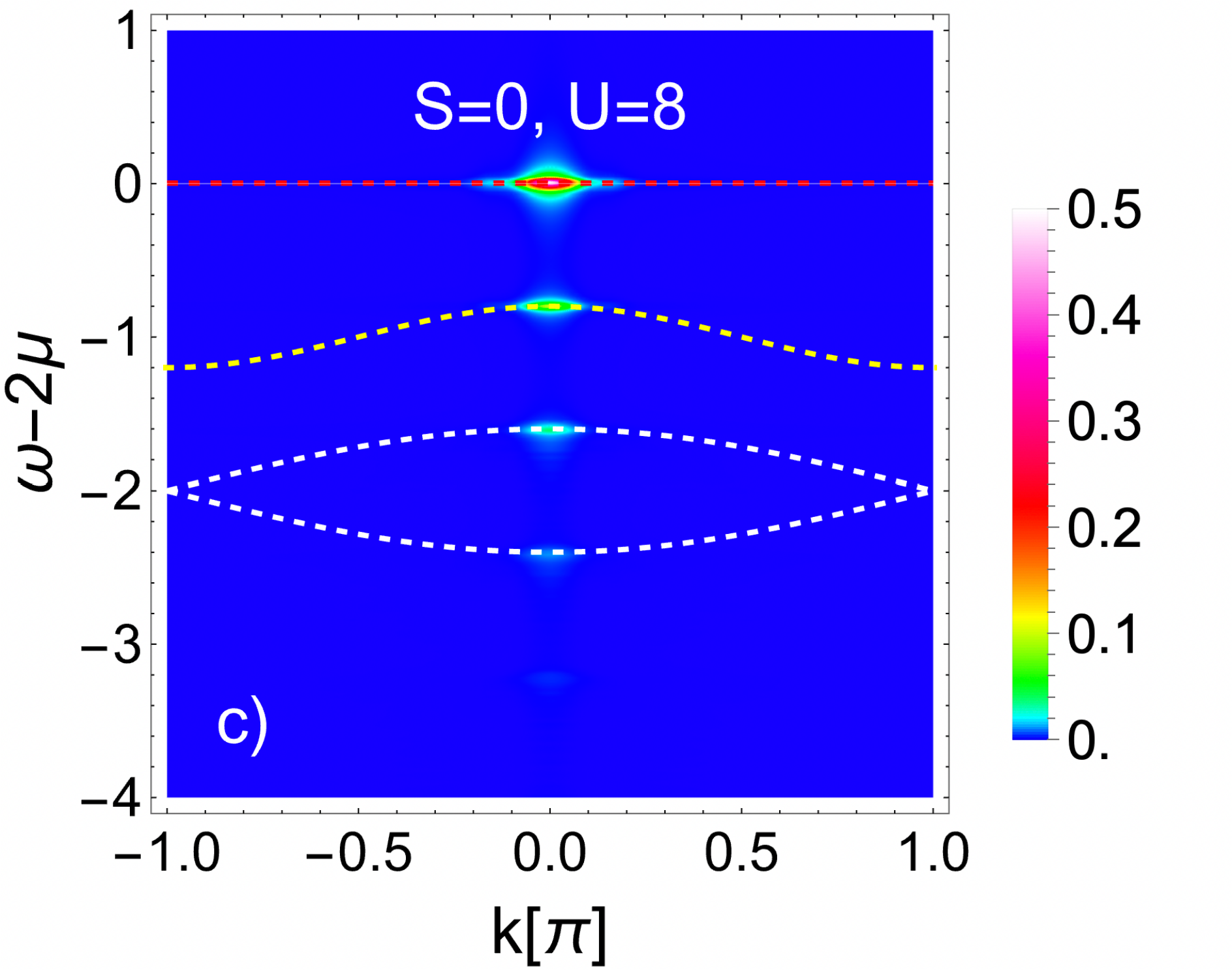}\includegraphics[width=0.5\columnwidth]{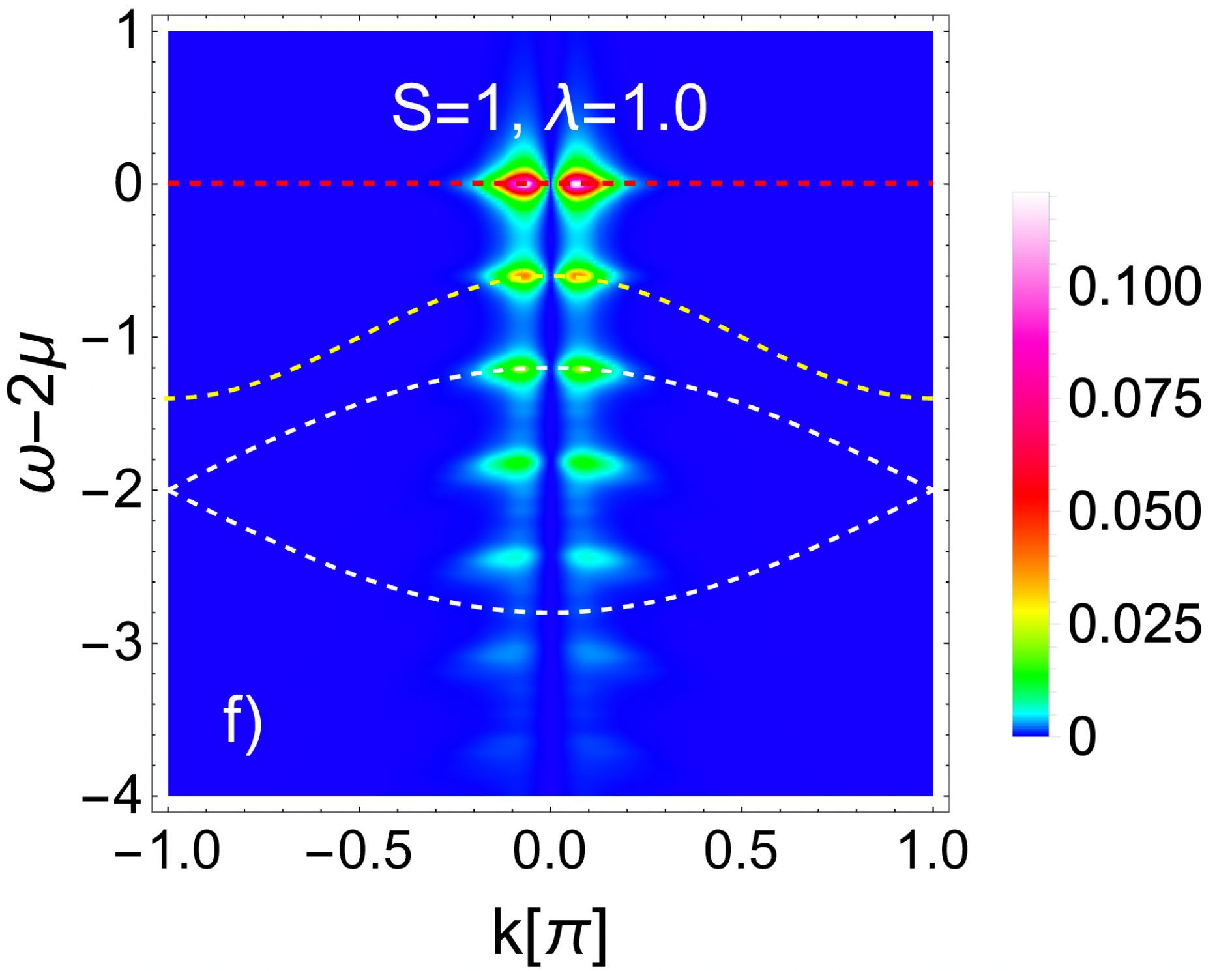}
\caption{  Contour plots of $A_2(\omega,-k,k)$ for $N_e=2$
and $\omega_0=1$. Panels a), b), c) are in the singlet channel with $\lambda=1, t_\mathrm{ph}=-0.1$ and $U=0,2,8$ respectively.  Panels  d), e) and f) are in the triplet channel with $t_\mathrm{ph}=-0.2$ and $\lambda = 1.75, 1.5, 1$ respectively. In all panels, the upper yellow dashed line shows  the single-phonon sideband at  $\omega_\mat{1ph}(k)=-\Omega_k$, and the two lower white dashed lines  enclose the shifted continuum representing 2--phonon excitations: $\omega_\mat{2ph}^\pm(k)=-2\omega_0 \pm 4t_\mat{ph} \cos(k/2)$. The red dashed line at $\omega - 2\mu = -\Delta$ marks the upper edge of the 2eARPES signal from electrons coming from different pairs (not shown). A thin white line indicates $\omega-2\mu=0$. 
 We used $N_\mat{h}=18$ and a system size  $N=64$.   }
\label{fig1n}
\end{figure}

Figure \ref{fig1n} shows contour plots of $A_2(\omega, k,-k)$ obtained with VED. We begin with panel (a) which shows the results in the singlet sector, when $U=0$.  Because of the strong el-ph coupling $\lambda =1$, the $N_e=2$ GS is a strongly bound bipolaron, hence  $A_2(\omega, k,-k)$ gives the 2e ARPES weight when both electrons are ejected from the same pair, with opposite momentum. As discussed above, the lowest binding energy feature appears indeed at $\omega-2\mu=0$ when no phonons are left behind. Conservation of momentum then requires $k_1+k_2=0$, explaining our choice $k_1=-k_2=k$ (we verified that this feature is invisible if $k_1+k_2\ne 0$). 

The next lowest binding energy feature has one phonon in the final state, with momentum $q=-k_1-k_2\rightarrow 0$ if $k_1=-k_2=k$. Indeed, its energy shift agrees with the phonon energy, shown by the yellow dashed line, at $q=0$. We checked that this feature is visible at other values of $k_1+k_2$ and follows the phonon dispersion (see Supp. Matt. \cite{SM}). The next lowest binding energy feature has two phonons with momenta $q_1+q_2=-k_1-k_2$ in the final state,  forming a continuum with weight expected in between the two lower white dashed lines, in agreement with the numerical results. The observation of these higher binding energy features allows the direct measurement of the phonon dispersion. It may also be possible to extract information about the  nature  of the electron-phonon coupling from their weight \cite{Krsnik2020}. However,  we remind the reader that in a system with a finite concentration of carriers, all features with energy $\omega-2\mu< -\Delta$ (horizontal red dashed line) are covered by the presumably much larger spectral weight from processes where the two electrons are ejected from different pairs, see Supp. Matt. for more details \cite{SM}.  In panel (a) both the one- and the two-phonon features are above this threshold, however as $U$ is increased and $\Delta $ decreases, eventually only the feature at $\omega - 2 \mu=0$ will be resolved (panel b). Finally, for a large $U$ (panel c), the $N_e=2$ GS corresponds to two unbound polarons and there is no  'single pair' signal.

Similar results are obtained for $A_2(\omega, k,-k)$ corresponding to emission from a triplet $p$-symmetry pair, as shown in panels d), e), f) of Fig. \ref{fig1n}. Here $U$ is irrelevant, and we can control the binding energy through varying $|\Delta E|$,  see discussion above. The obvious difference between the triplet and singlet pairs 2eARPES intensity is the 'node' obtained at $k=0$ for the former, as required by Pauli's principle.

Next, we analyze the evolution of the lowest binding energy feature at $\omega - 2 \mu=0$ with the binding energy, and argue that its existence (and thus, the existence of pre-formed pairs) can be confirmed even when the energy resolution is comparable or even worse than the gap size $\Delta$. Comparison of this feature in panels a) and b) of Fig. \ref{fig1n} shows that it becomes narrower in $k$-space as $\Delta $ decreases. This is easy to understand if we consider the bipolaron GS:
 \begin{align}   
& \vert\psi_0^{(GS,2)}\rangle = \sum_k \alpha_k c^\dagger_{k,\uparrow}c^\dagger_{-k,\downarrow} \vert 
   \emptyset\rangle + \sum_{k,q} \alpha_{k,q} c^\dagger_{k,\uparrow}c^\dagger_{-k-q,\downarrow} b^\dagger_q\vert  \emptyset\rangle \nonumber \\
  &  + \sum_{k,q_1,q_2} \alpha_{k,q_1,q_2} c^\dagger_{k,\uparrow}c^\dagger_{-k-q_1-q_2,\downarrow} b^\dagger_{q_1}b^\dagger_{q_2} \vert 
\emptyset\rangle+ \dots\nonumber
\end{align}

The spectral weight of the lowest binding energy feature (due to ejection of the two electrons and no phonons left behind) is proportional to $|\alpha_k|^2$. The Fourier transform of $\alpha_k$ defines the amplitude of probability to find the two bound electrons at a relative distance $\delta$ apart. For more strongly bound pairs, this peaks at a smaller $\delta$, implying a larger spread in $k$ -- precisely what is observed by comparing panels a) and b) for the singlet pair, and d) and e) for the triplet pair.

\begin{figure}[t]
\includegraphics[width=0.48\columnwidth]{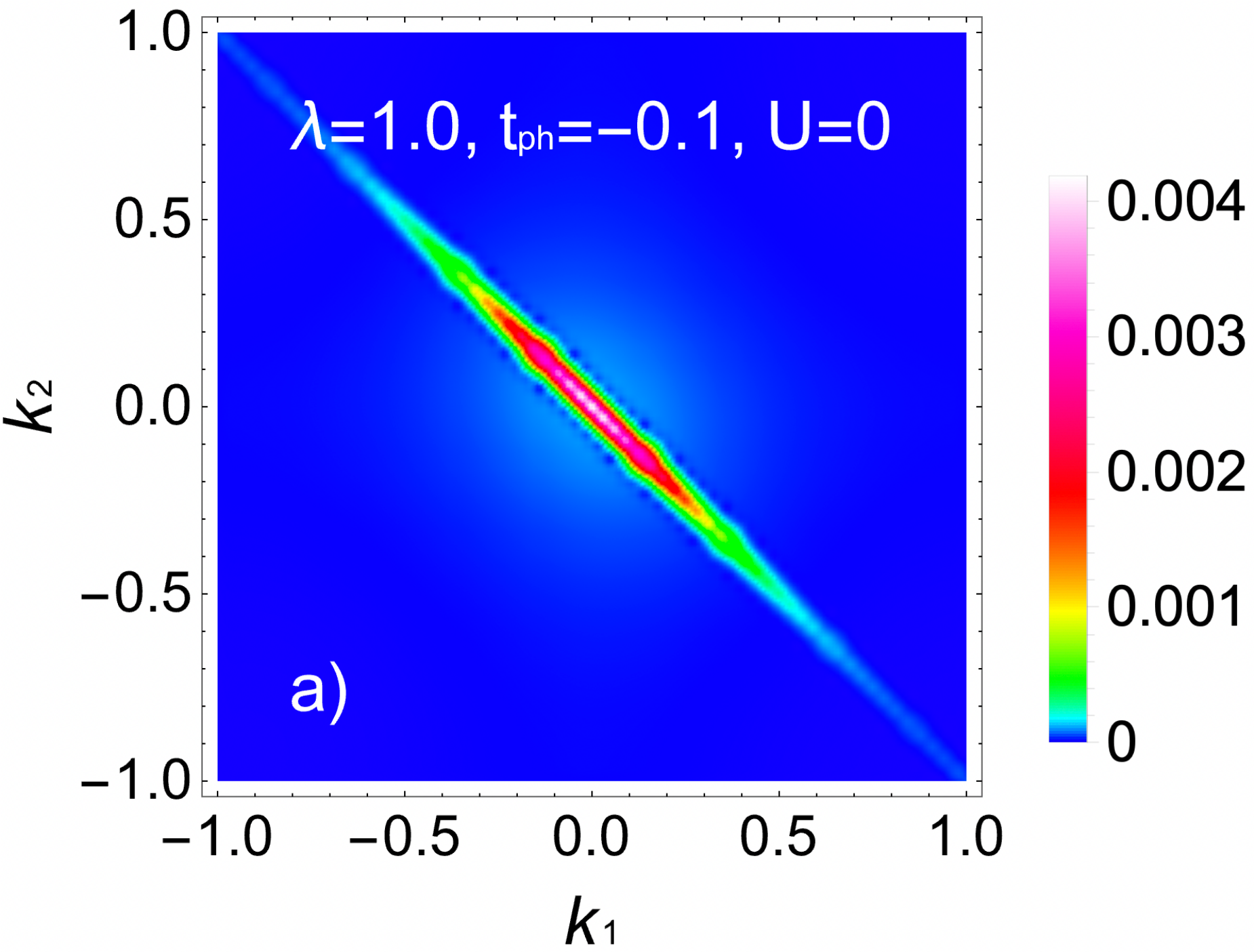}\includegraphics[width=0.5\columnwidth]{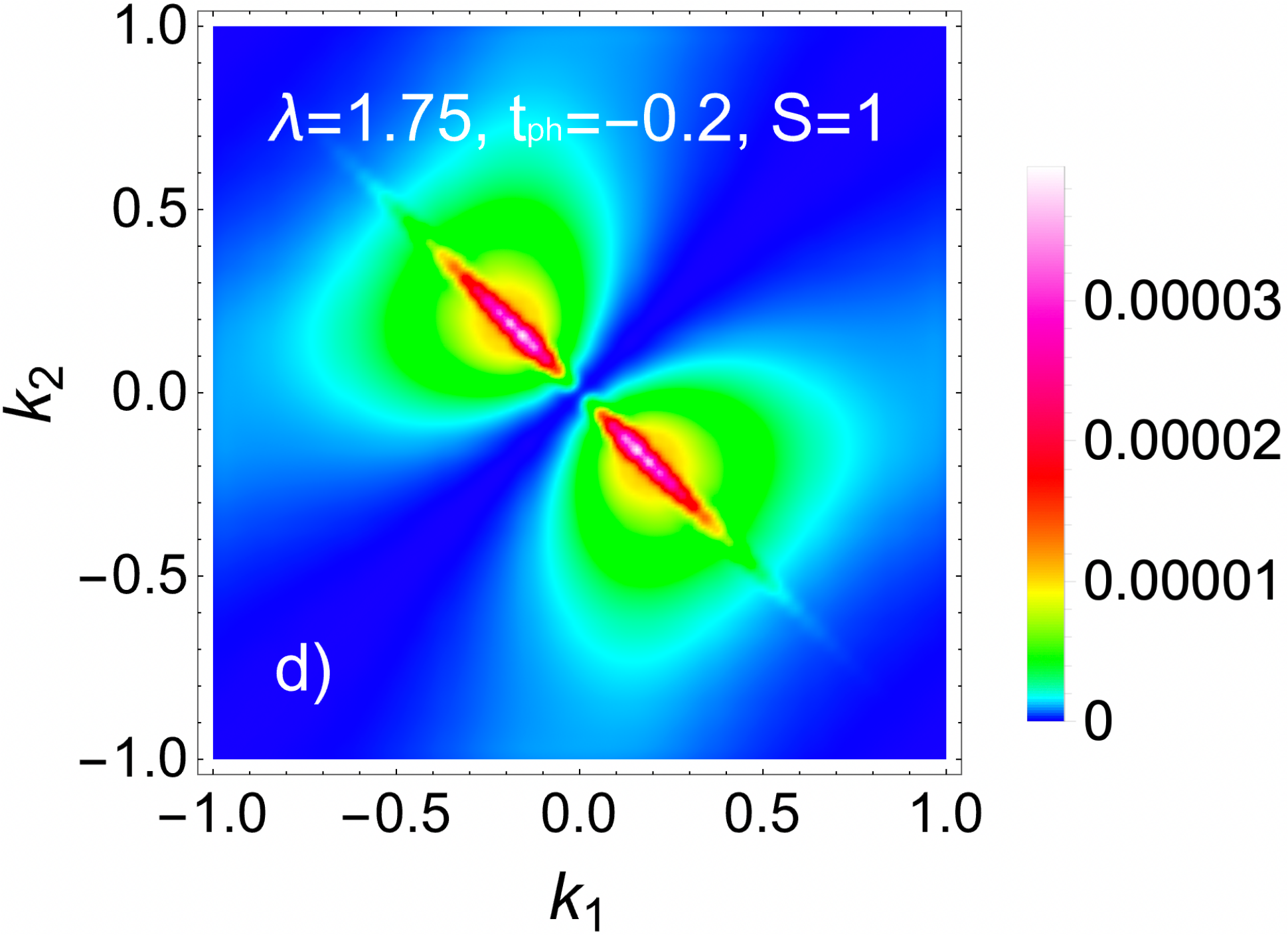}
\includegraphics[width=0.48\columnwidth]{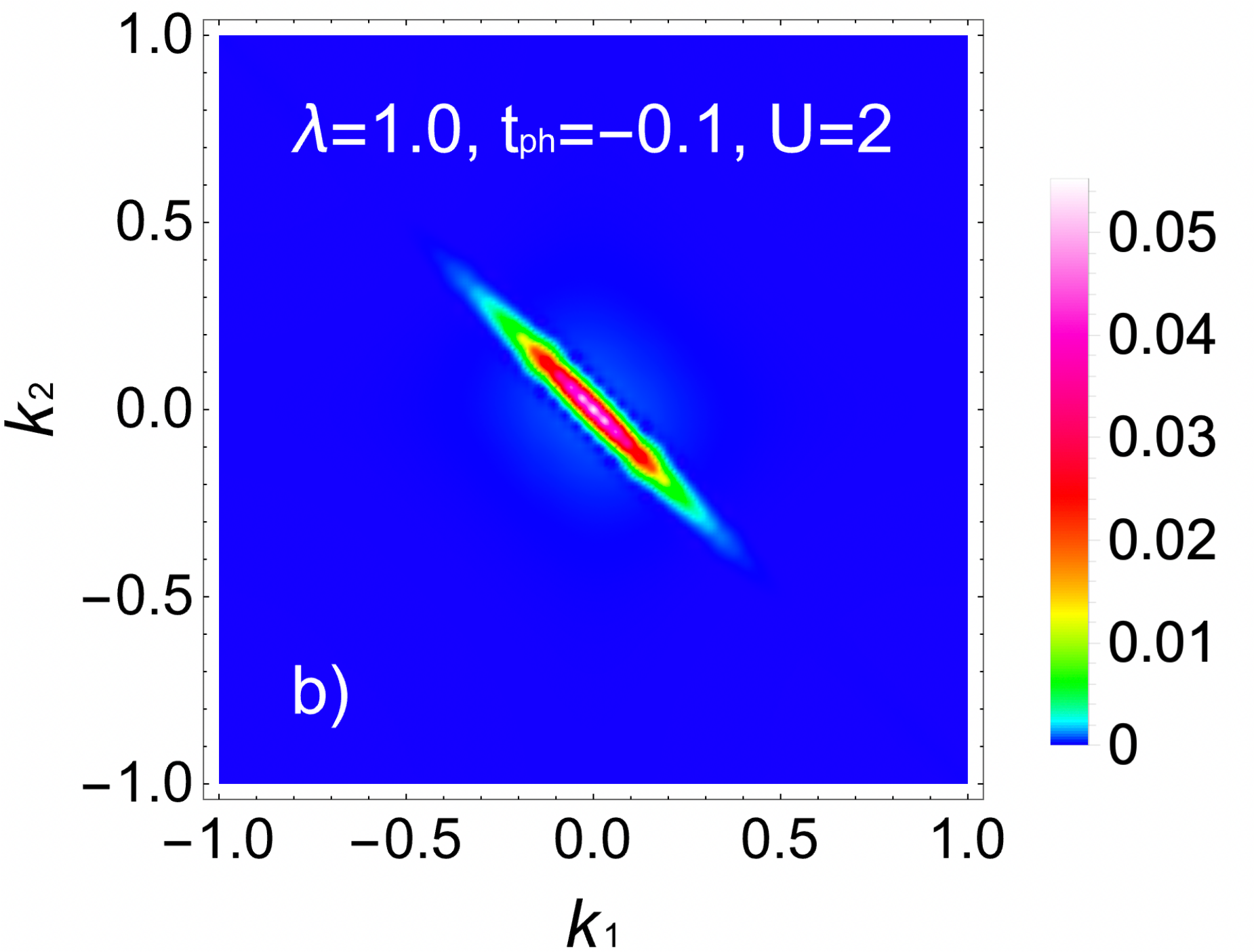}\includegraphics[width=0.5\columnwidth]{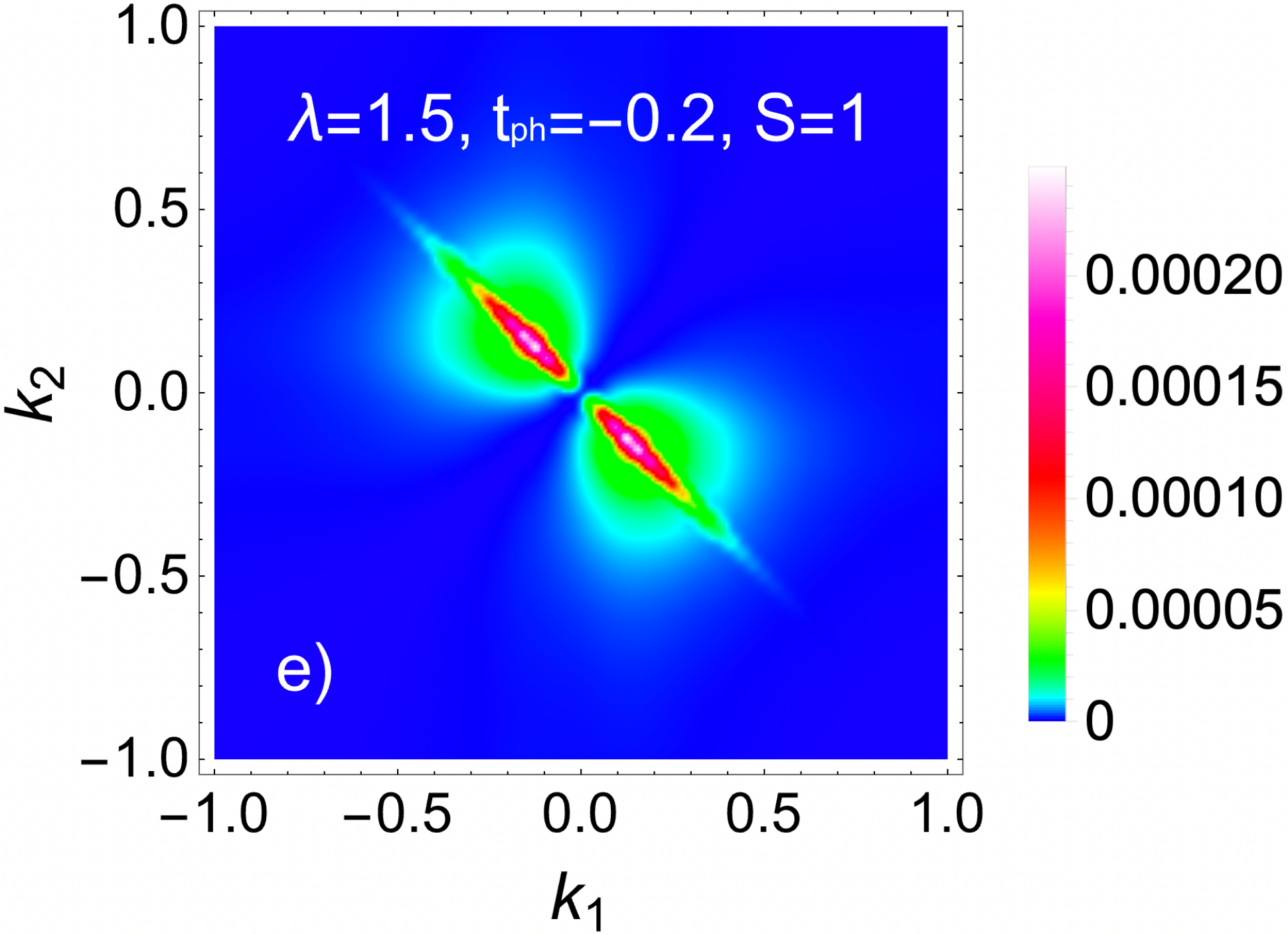}
\includegraphics[width=0.48\columnwidth]{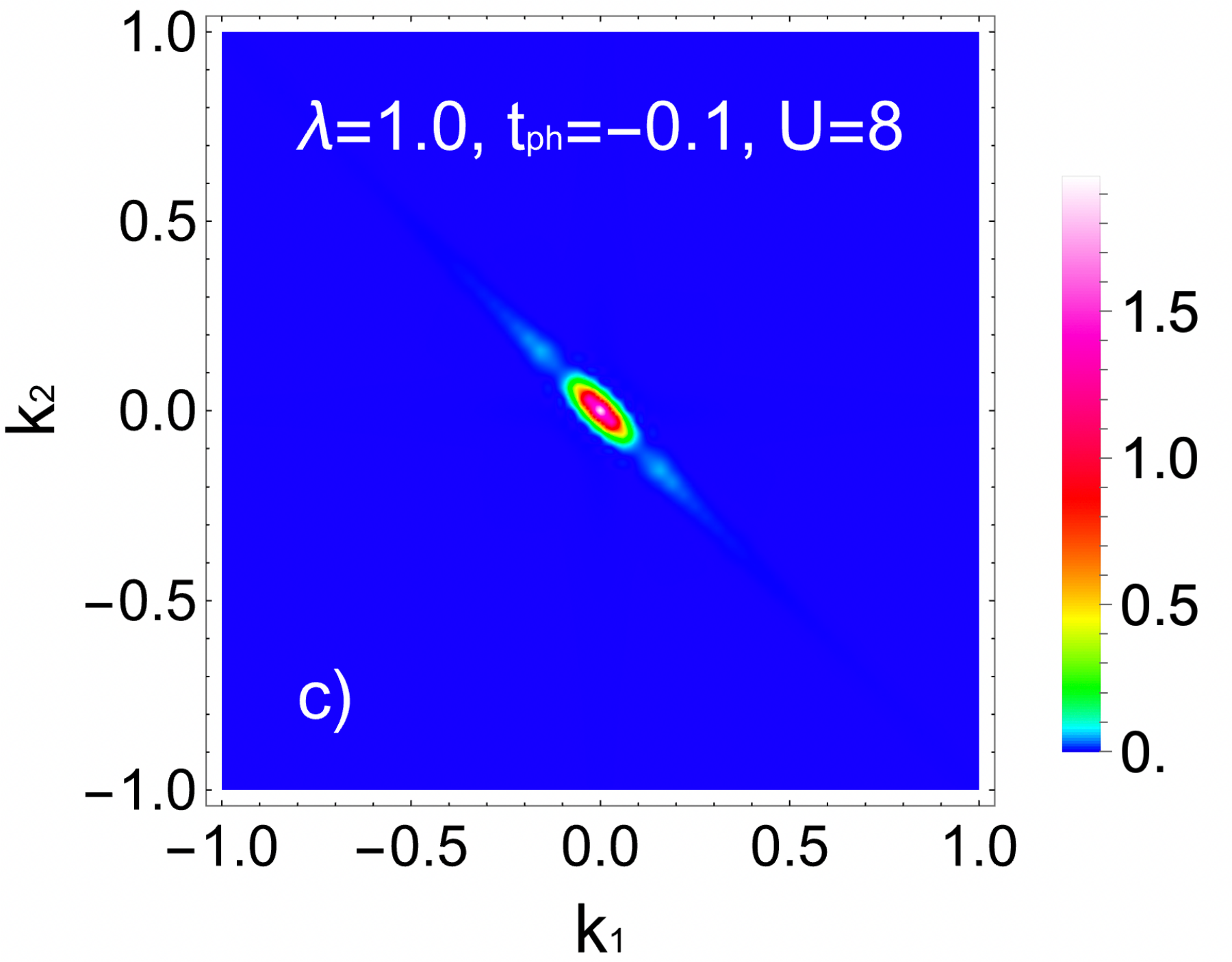}\includegraphics[width=0.5\columnwidth]{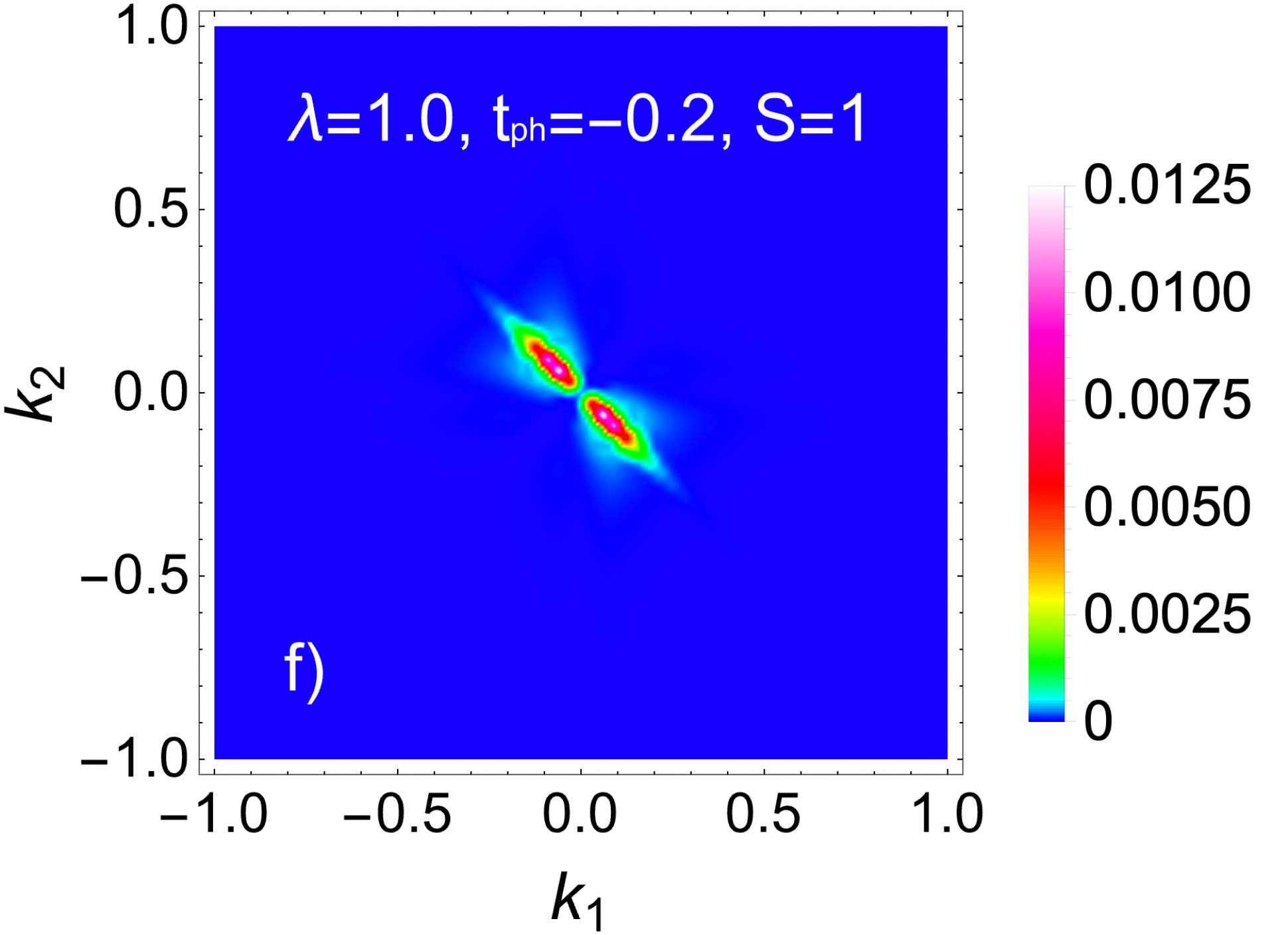}
\caption{   Momentum-resolved spectral weight of the lowest-binding energy feature $\gamma_{pair}(k_1,k_2)= A_2(\omega=2\mu, k_1, k_2 )$.  In order to increase the number of $k-$ points, we used  $N_\mathrm{h}=16$ and $N=64$.}
\label{fig2n}
\end{figure}

Another way to illustrate this is to show the momentum resolved 2eARPES weight $\gamma_{pair}(k_1,k_2)= A_2(\omega=2\mu, k_1, k_2 )$ due to this feature only, which can be obtained by integrating over a narrow energy range. Figure \ref{fig2n} shows $\gamma_{pair}(k_1,k_2)$ for parameters identical to those in Fig. \ref{fig1n},  illustrating beautifully the inverse relationship between the spread of this feature along the $k_1+k_2=0$ line, and the binding energy of the pair.

These results verify that the 2eARPES signal at $\omega = 2 \mu$  that originates when the electrons are emitted from the same pair appears only when $k_1=-k_2$.  If the detectors are at equal distance from the sample and select photoelectrons with equal momenta $|{\bf k}_1|= |{\bf k}_2|$, then the times of flight are equal.
  Coincidence detection of such photoelectrons means that they must have been emitted at the same time {\em i.e.} $\tau =0$, explaining why we made this choice when simplifying Eq. (\ref{Dev}). However, the conservation of momentum of the system which underlies the $k_1+k_2=0$ condition is always valid, hence we expect a similar conclusion even if the full Eq. (\ref{Dev}) is evaluated.

This asymmetry of the single-pair signal in the $(k_1,k_2)$ plane explains how the existence of pre-formed pairs could be inferred even if the binding energy $\Delta$ was smaller than the energy resolution of the apparatus, so that  this signal was covered by the presumably larger signal from electrons emitted from different pairs. Based on Eq. (2), we can estimate the latter to be proportional to the convolution  $\int d\omega_1 \int d\omega_2  A_1(\omega_1, k_1) A_1(\omega_2, k_2)\delta(\omega - \omega_1-\omega_2)$, where the (ARPES) single particle spectral weights are:
\begin{equation}
    A_1(\omega, k) = {1\over \pi} \mbox{Im} \sum_{n} \frac{|\langle \psi^{(n,N_e-1)}_{ -k}| c_{k\sigma}| \psi^{(GS,N_e)}_0\rangle|^2}{\omega - i \eta +E_{-k}^{(n,N_e-1)}-E_0^{(GS,N_e)}}
\end{equation}
Let $\beta(k_1,k_2)$ be the result when this signal from two different pairs is integrated for a  small energy range below $\omega=2\mu-\Delta$. Because $A_1(\omega,k) = A_1(\omega,-k) $, it follows immediately that $\beta(k_1,k_2)= \beta(- k_1, k_2)=\beta(k_1,-k_2)= \beta(- k_1, -k_2)$, {\em i.e.} this countour plot has $C_4$ symmetry in the $(k_1,k_2)$ space, unlike the strong $C_2$ symmetry of $\gamma_{pair}(k_1,k_2)$ (plots of $\beta(k_1,k_2)$ are shown in the Supplementary Material \cite{SM}). As a result, if the total energy-integrated signal at the top of the spectrum has more weight along  $k_1+k_2=0$  than along $k_1=k_2$, this confirms the existence of pre-formed pairs.

While the numerical results presented here are at $T=0$ and for a single pair on a 1D chain, we believe that these two fingerprints pointing to the existence of pre-formed pairs are present at finite-$T$ for finite densities in any dimension. 

At finite temperatures, a fraction of the pairs are thermally dissociated. As noted above, 2eARPES signal from unbound quasiparticles starts at $\omega-2\mu=0$, so it is always superimposed over the 2eARPES signal coming from electrons ejected from the same pair. Nevertheless, its $C_4$ symmetry (and different evolution with temperature) should make it distinguishable from the pair signal. 

At finite concentrations and low-$T$, the system is superconducting if a macroscopic fraction of the
the pairs condense in the $K=0$ state. In this case, the 2eARPES signal coming from electrons ejected from the same pair is proportional to that shown in Figs. \ref{fig1n} and \ref{fig2n} for a single pair with $K=0$,  in agreement with the findings of  Ref. \cite{Devereaux_2023}. 

Another option is an incoherent liquid of pre-formed pairs,  where each pair momentum $K$ has a microscopic occupation number $n_K/N\rightarrow 0$. This occurs at $T=0$  for our 1D model \cite{Klemen2025}, but more generally might be expected as an intermediate state between a superconductor and the normal state, if superconductivity is lost due to phase fluctuations (not to un-pairing, like in BCS). The occupied pair $K$ momenta are set by $k_F$, as demonstrated in Ref. \cite{Klemen2025} for the 'Bose sea' GS of our model. The 2eARPES signal from electrons ejected from a pair with momentum $K\ne 0$ is similar to those shown in Fig. \ref{fig2n} but shifted to $k_1+k_2=K$. Thus, the very asymmetric $C_2$ pattern remains, however it acquires a finite 'transverse' width proportional to $k_F$, as further discussed in the End Matter.  (In  Figs. \ref{fig1n},\ref{fig2n}, the widths are set by the  broadening $\eta$). All these considerations carry over in higher-D in the plane parallel to the sample surface,  ${\bf k}_{1,\parallel}+{\bf k}_{2,\parallel}={\bf K}_{\parallel}$.

{\em Summary:}  To conclude, we identified two fingerprints that allow the separation of the presumably small contribution of processes where both electrons are emitted from the same pair, from the total 2eARPES intensity. These fingerprints are direct consequences of the conservation of momentum and energy, therefore we are confident that they will also appear in more sophisticated theoretical descriptions of 2eARPES spectroscopy and in more comprehensive numerical methods, such as Diagrammatic Monte Carlo, which can calculate 2eARPES intensities for finite concentrations in the thermodynamic limit of Hamiltonian (\ref{ham}) in any dimension \cite{Burovski2008,Mishchenko2014,Tupitsyn2016,Mishchenko2018}. Experimental observation of these fingerprints will confirm the existence of electron pairs in the system, and whether they are coherent (superconducting) or not.  

{\em Acknowledgements:} J.B. acknowledges the support by the program No. P1-0044 and   VIP project KTTK21 under contract no. SN-ZRD/22-27/510 of the Slovenian Research and Innovation Agency (ARIS). J.B.  acknowledges  discussions with S.A. Trugman, A. Saxena and support from  the Center for Integrated Nanotechnologies, a U.S. Department of Energy, Office of Basic Energy Sciences user facility and Quantum and Condensed Matter Physics  (T-4) at Los Alamos National Laboratory. This project was undertaken thanks in part to funding from the Max Planck-UBC-UTokyo Center for
Quantum Materials and the Canada First Research Excellence Fund, Quantum Materials and Future Technologies Program, as well as the  Natural Sciences and Engineering Research Council of Canada (A.D. and M. B.), the Canada Research Chairs Program (A.D.) and the CIFAR Quantum Materials Program (A.D.).

\appendix

\section{End matter}

\begin{figure}[t]
\includegraphics[width=0.48\columnwidth]{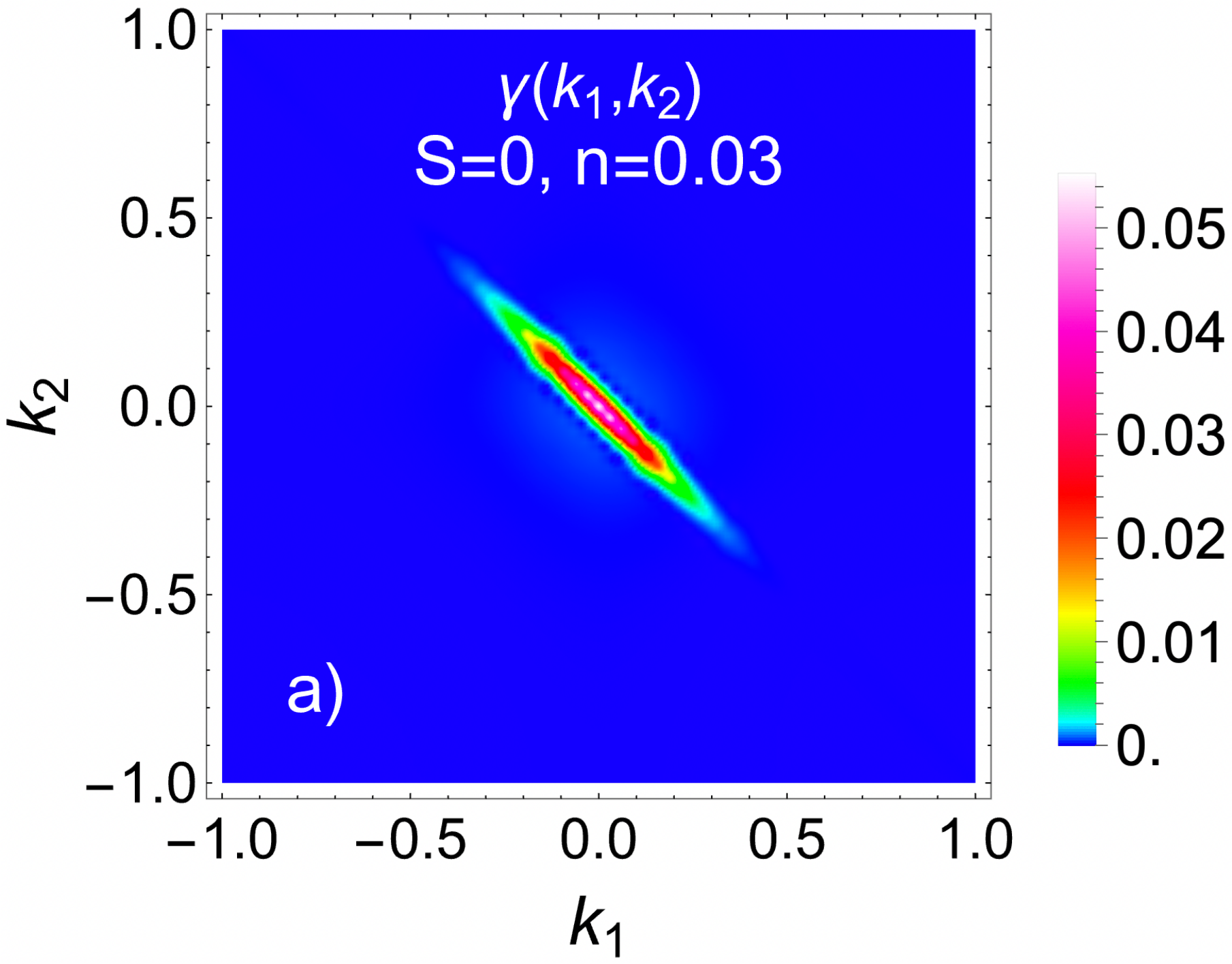}\includegraphics[width=0.5\columnwidth]{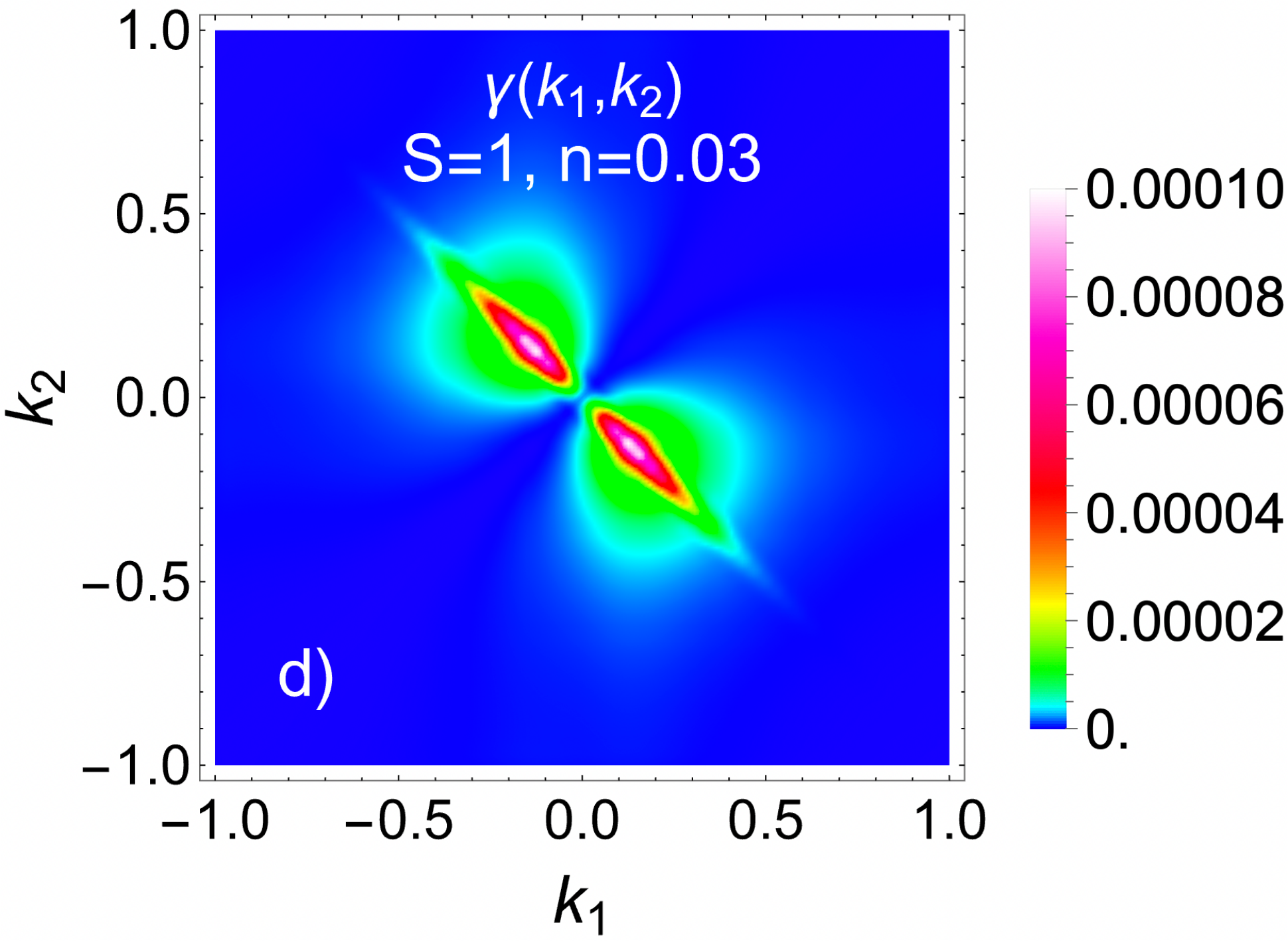}
\includegraphics[width=0.48\columnwidth]{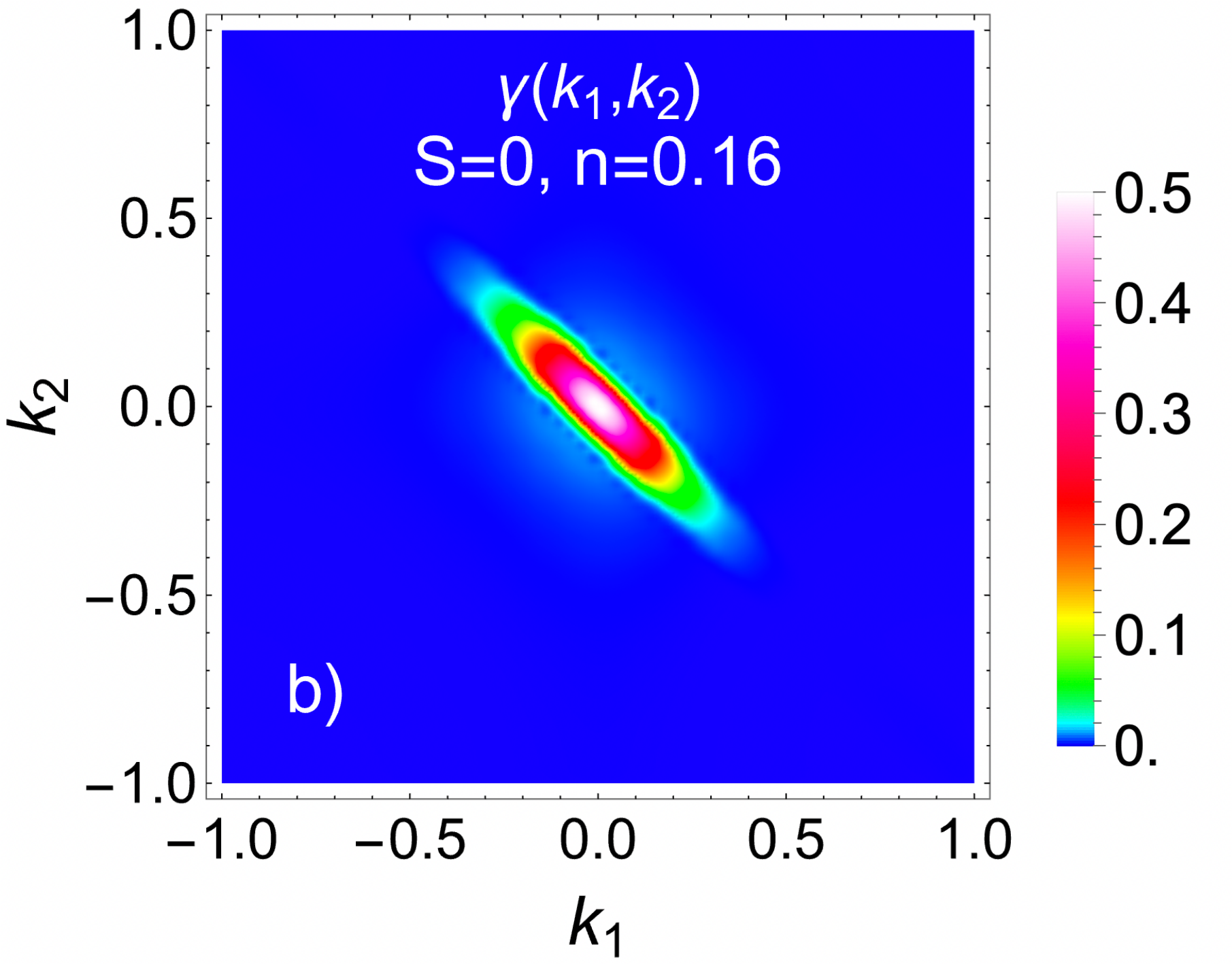}\includegraphics[width=0.5\columnwidth]{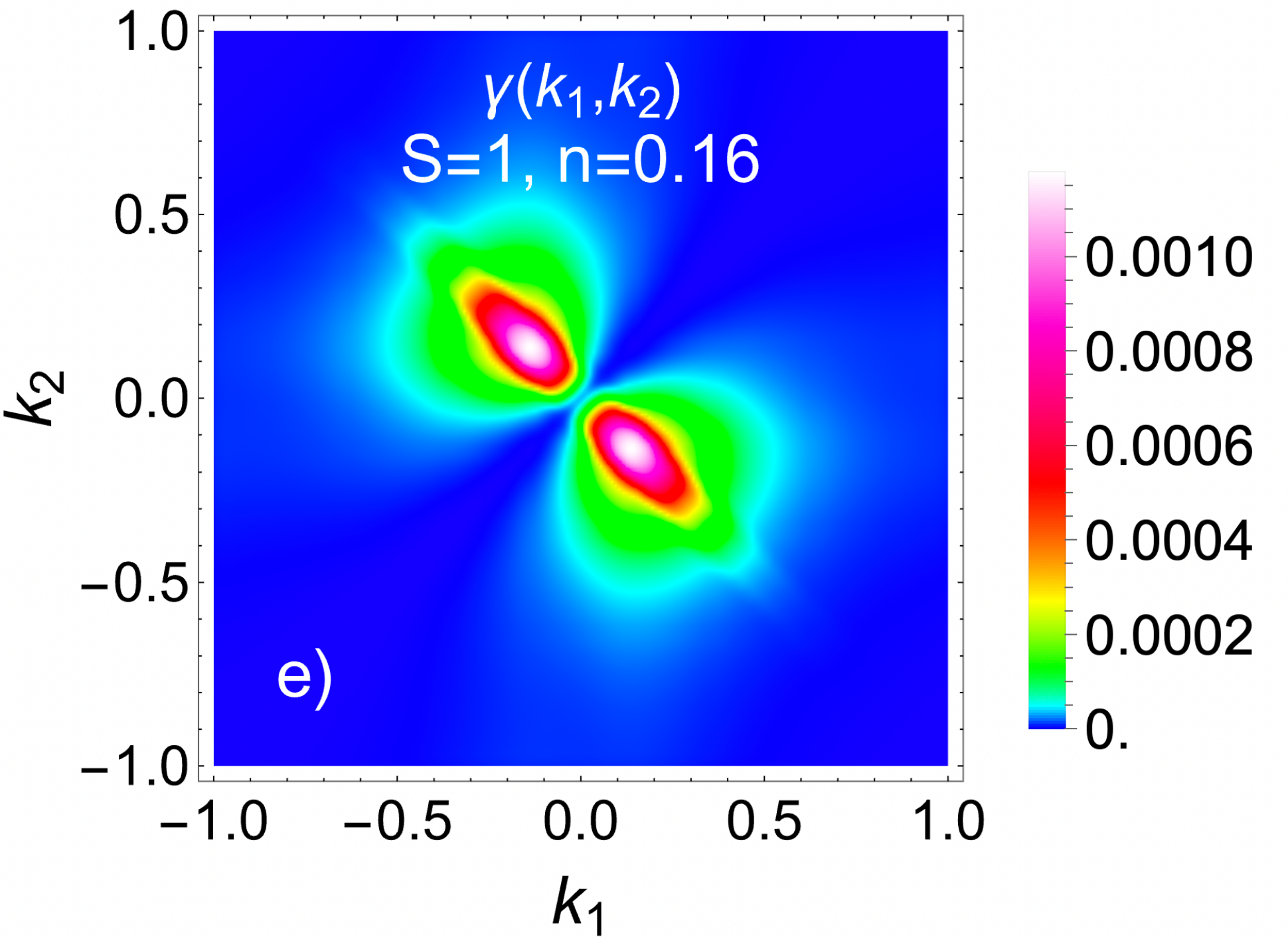}
\includegraphics[width=0.48\columnwidth]{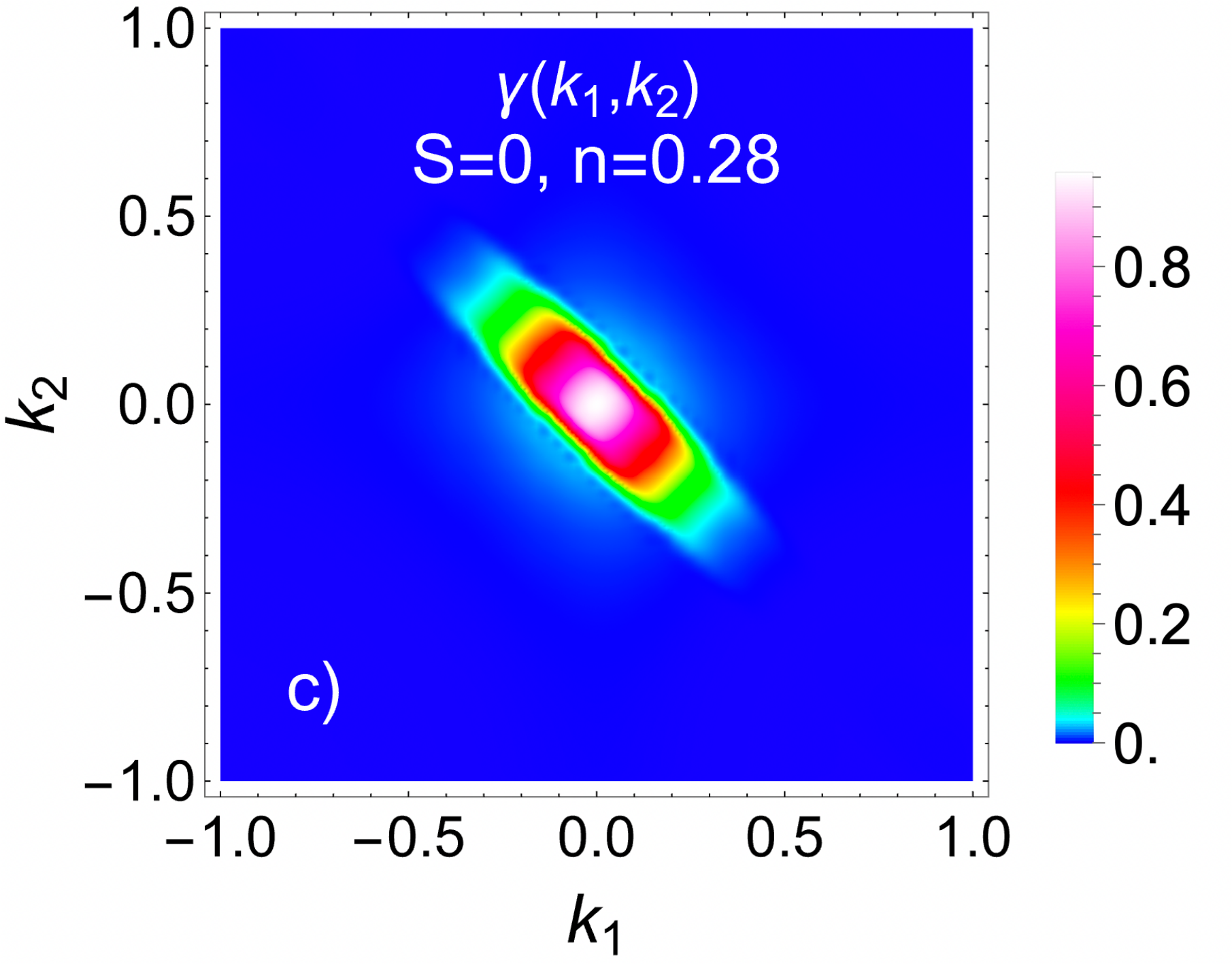}\includegraphics[width=0.5\columnwidth]{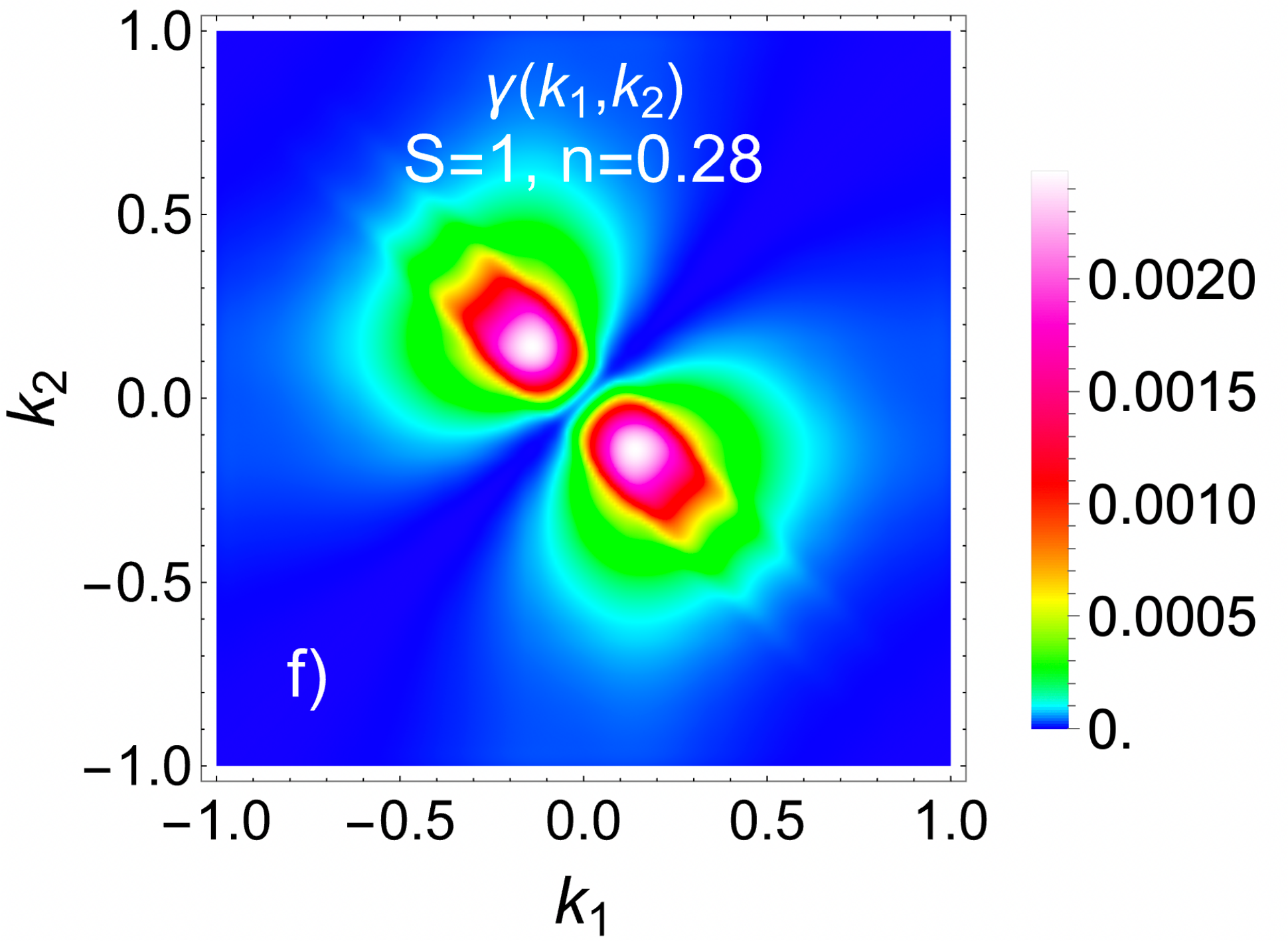}
\caption{   Momentum-resolved spectral weight of the lowest-binding energy feature $\gamma_{pair}(k_1,k_2)= A_2(\omega=2\mu, k_1, k_2 )$ for densities $n=N_e/N=0.03, 0.16, 0.28$ in the top, middle and bottom panels, respectively. The left column panels are in the singlet channel with $\lambda= 1, t_{\rm ph}=-0.1, U=2$, while the right column panels are in the triplet channel with $\lambda= 1.75, t_{\rm ph}=-0.2$. }
\label{fig3n}
\end{figure}

Figure \ref{fig3n} shows the evolution of the momentum-resolved spectral weight of the lowest-binding energy feature $\gamma_{pair}(k_1,k_2)= A_2(\omega=2\mu, k_1, k_2 )$ (integrated over a narrow energy range) with increasing electron density $n=N_e/N=0.03, 0.16$ and 0.28 (from top to bottom panels). The left column panels are for the singlet channel with $\lambda= 1, t_{\rm ph}=-0.1, U=2$, while the right column panels are in the triplet channel with $\lambda= 1.75, t_{\rm ph}=-0.2$. For these values, the GS are liquids of incoherent pairs (coherence is not possible in 1D because of quantum fluctuations). 

These results were obtained by summing individual 2eARPES intensities obtained with VED from Eq. (\ref{G}) for individual pairs with momenta  $|K|\le n\pi=k_F$. This 'Bose sea' of non-interacting pairs was shown in Ref. \cite{Klemen2025} to reproduce accurately results obtained with density matrix renormalization group (DMRG) at finite concentrations.

The results of Fig. \ref{fig3n} confirm that even for a finite-density incoherent liquid of pairs, the strong $C_2$ symmetry is maintained. This is a direct consequence of momentum conservation, as the two electrons ejected from a pair with momentum $K$ must obey $k_1+k_2=K$. The 'transverse' broadening then gives a direct measure of the occupied pair momenta $K$.


\begin{thebibliography}{22}
\expandafter\ifx\csname natexlab\endcsname\relax\def\natexlab#1{#1}\fi
\expandafter\ifx\csname bibnamefont\endcsname\relax
  \def\bibnamefont#1{#1}\fi
\expandafter\ifx\csname bibfnamefont\endcsname\relax
  \def\bibfnamefont#1{#1}\fi
\expandafter\ifx\csname citenamefont\endcsname\relax
  \def\citenamefont#1{#1}\fi
\expandafter\ifx\csname url\endcsname\relax
  \def\url#1{\texttt{#1}}\fi
\expandafter\ifx\csname urlprefix\endcsname\relax\def\urlprefix{URL }\fi
\providecommand{\bibinfo}[2]{#2}
\providecommand{\eprint}[2][]{\url{#2}}

\bibitem[{\citenamefont{Damascelli et~al.}(2003)\citenamefont{Damascelli, Hussain, and Shen}}]{Damascelli2003}
\bibinfo{author}{\bibfnamefont{A.}~\bibnamefont{Damascelli}}, \bibinfo{author}{\bibfnamefont{Z.}~\bibnamefont{Hussain}}, \bibnamefont{and} \bibinfo{author}{\bibfnamefont{Z.-X.} \bibnamefont{Shen}}, \bibinfo{journal}{Rev. Mod. Phys.} \textbf{\bibinfo{volume}{75}}, \bibinfo{pages}{473} (\bibinfo{year}{2003}), \urlprefix\url{https://link.aps.org/doi/10.1103/RevModPhys.75.473}.

\bibitem[{\citenamefont{Damascelli}(2004)}]{Damascelli2004}
\bibinfo{author}{\bibfnamefont{A.}~\bibnamefont{Damascelli}}, \bibinfo{journal}{Physica Scripta} \textbf{\bibinfo{volume}{2004}}, \bibinfo{pages}{61} (\bibinfo{year}{2004}), \urlprefix\url{https://dx.doi.org/10.1238/Physica.Topical.109a00061}.

\bibitem[{\citenamefont{Kova\ifmmode~\check{c}\else \v{c}\fi{} et~al.}(2025)\citenamefont{Kova\ifmmode~\check{c}\else \v{c}\fi{}, Nocera, Damascelli, Bon\ifmmode~\check{c}\else \v{c}\fi{}a, and Berciu}}]{Klemen2025}
\bibinfo{author}{\bibfnamefont{K.}~\bibnamefont{Kova\ifmmode~\check{c}\else \v{c}\fi{}}}, \bibinfo{author}{\bibfnamefont{A.}~\bibnamefont{Nocera}}, \bibinfo{author}{\bibfnamefont{A.}~\bibnamefont{Damascelli}}, \bibinfo{author}{\bibfnamefont{J.}~\bibnamefont{Bon\ifmmode~\check{c}\else \v{c}\fi{}a}}, \bibnamefont{and} \bibinfo{author}{\bibfnamefont{M.}~\bibnamefont{Berciu}}, \bibinfo{journal}{Phys. Rev. Lett.} \textbf{\bibinfo{volume}{134}}, \bibinfo{pages}{096502} (\bibinfo{year}{2025}), \urlprefix\url{https://link.aps.org/doi/10.1103/PhysRevLett.134.096502}.

\bibitem[{\citenamefont{Su et~al.}(2025)\citenamefont{Su, Zhang, Zhang, and Cao}}]{Su2025}
\bibinfo{author}{\bibfnamefont{Y.}~\bibnamefont{Su}}, \bibinfo{author}{\bibfnamefont{G.}~\bibnamefont{Zhang}}, \bibinfo{author}{\bibfnamefont{C.}~\bibnamefont{Zhang}}, \bibnamefont{and} \bibinfo{author}{\bibfnamefont{D.}~\bibnamefont{Cao}}, \emph{\bibinfo{title}{Coincidence detection techniques for direct measurement of many-body correlations in strongly correlated electron systems}} (\bibinfo{year}{2025}), \eprint{2512.06593}, \urlprefix\url{https://arxiv.org/abs/2512.06593}.

\bibitem[{\citenamefont{Schumann et~al.}(2007)\citenamefont{Schumann, Winkler, and Kirschner}}]{Kirschner_2007}
\bibinfo{author}{\bibfnamefont{F.~O.} \bibnamefont{Schumann}}, \bibinfo{author}{\bibfnamefont{C.}~\bibnamefont{Winkler}}, \bibnamefont{and} \bibinfo{author}{\bibfnamefont{J.}~\bibnamefont{Kirschner}}, \bibinfo{journal}{Phys. Rev. Lett.} \textbf{\bibinfo{volume}{98}}, \bibinfo{pages}{257604} (\bibinfo{year}{2007}), \urlprefix\url{https://link.aps.org/doi/10.1103/PhysRevLett.98.257604}.

\bibitem[{\citenamefont{Mahmood et~al.}(2022)\citenamefont{Mahmood, Devereaux, Abbamonte, and Morr}}]{T2}
\bibinfo{author}{\bibfnamefont{F.}~\bibnamefont{Mahmood}}, \bibinfo{author}{\bibfnamefont{T.}~\bibnamefont{Devereaux}}, \bibinfo{author}{\bibfnamefont{P.}~\bibnamefont{Abbamonte}}, \bibnamefont{and} \bibinfo{author}{\bibfnamefont{D.~K.} \bibnamefont{Morr}}, \bibinfo{journal}{Phys. Rev. B} \textbf{\bibinfo{volume}{105}}, \bibinfo{pages}{064515} (\bibinfo{year}{2022}), \urlprefix\url{https://link.aps.org/doi/10.1103/PhysRevB.105.064515}.

\bibitem[{\citenamefont{Kemper et~al.}(2025)\citenamefont{Kemper, Goto, Labib, Gauthier, da~Silva~Neto, and Boschini}}]{Kemper2025}
\bibinfo{author}{\bibfnamefont{A.~F.} \bibnamefont{Kemper}}, \bibinfo{author}{\bibfnamefont{F.}~\bibnamefont{Goto}}, \bibinfo{author}{\bibfnamefont{H.~A.} \bibnamefont{Labib}}, \bibinfo{author}{\bibfnamefont{N.}~\bibnamefont{Gauthier}}, \bibinfo{author}{\bibfnamefont{E.~H.} \bibnamefont{da~Silva~Neto}}, \bibnamefont{and} \bibinfo{author}{\bibfnamefont{F.}~\bibnamefont{Boschini}}, \emph{\bibinfo{title}{Observing two-electron interactions with correlation-arpes}} (\bibinfo{year}{2025}), \eprint{2505.01504}, \urlprefix\url{https://arxiv.org/abs/2505.01504}.

\bibitem[{\citenamefont{Devereaux et~al.}(2023)\citenamefont{Devereaux, Claassen, Huang, Zaletel, Moore, Morr, Mahmood, Abbamonte, and Shen}}]{Devereaux_2023}
\bibinfo{author}{\bibfnamefont{T.~P.} \bibnamefont{Devereaux}}, \bibinfo{author}{\bibfnamefont{M.}~\bibnamefont{Claassen}}, \bibinfo{author}{\bibfnamefont{X.-X.} \bibnamefont{Huang}}, \bibinfo{author}{\bibfnamefont{M.}~\bibnamefont{Zaletel}}, \bibinfo{author}{\bibfnamefont{J.~E.} \bibnamefont{Moore}}, \bibinfo{author}{\bibfnamefont{D.}~\bibnamefont{Morr}}, \bibinfo{author}{\bibfnamefont{F.}~\bibnamefont{Mahmood}}, \bibinfo{author}{\bibfnamefont{P.}~\bibnamefont{Abbamonte}}, \bibnamefont{and} \bibinfo{author}{\bibfnamefont{Z.-X.} \bibnamefont{Shen}}, \bibinfo{journal}{Phys. Rev. B} \textbf{\bibinfo{volume}{108}}, \bibinfo{pages}{165134} (\bibinfo{year}{2023}), \urlprefix\url{https://link.aps.org/doi/10.1103/PhysRevB.108.165134}.

\bibitem[{\citenamefont{Stahl and Eckstein}(2019)}]{Eckstein2019}
\bibinfo{author}{\bibfnamefont{C.}~\bibnamefont{Stahl}} \bibnamefont{and} \bibinfo{author}{\bibfnamefont{M.}~\bibnamefont{Eckstein}}, \bibinfo{journal}{Phys. Rev. B} \textbf{\bibinfo{volume}{99}}, \bibinfo{pages}{241111} (\bibinfo{year}{2019}), \urlprefix\url{https://link.aps.org/doi/10.1103/PhysRevB.99.241111}.

\bibitem[{\citenamefont{Su and Zhang}(2020)}]{Su2020}
\bibinfo{author}{\bibfnamefont{Y.}~\bibnamefont{Su}} \bibnamefont{and} \bibinfo{author}{\bibfnamefont{C.}~\bibnamefont{Zhang}}, \bibinfo{journal}{Phys. Rev. B} \textbf{\bibinfo{volume}{101}}, \bibinfo{pages}{205110} (\bibinfo{year}{2020}), \urlprefix\url{https://link.aps.org/doi/10.1103/PhysRevB.101.205110}.

\bibitem[{\citenamefont{Marchand and Berciu}(2013)}]{berciu2013}
\bibinfo{author}{\bibfnamefont{D.~J.~J.} \bibnamefont{Marchand}} \bibnamefont{and} \bibinfo{author}{\bibfnamefont{M.}~\bibnamefont{Berciu}}, \bibinfo{journal}{Phys. Rev. B} \textbf{\bibinfo{volume}{88}}, \bibinfo{pages}{060301} (\bibinfo{year}{2013}), \urlprefix\url{https://link.aps.org/doi/10.1103/PhysRevB.88.060301}.

\bibitem[{\citenamefont{Bon\v{c}a et~al.}(1999)\citenamefont{Bon\v{c}a, Trugman, and Batisti\'{c}}}]{bonca1}
\bibinfo{author}{\bibfnamefont{J.}~\bibnamefont{Bon\v{c}a}}, \bibinfo{author}{\bibfnamefont{S.~A.} \bibnamefont{Trugman}}, \bibnamefont{and} \bibinfo{author}{\bibfnamefont{I.}~\bibnamefont{Batisti\'{c}}}, \bibinfo{journal}{Phys. Rev. B} \textbf{\bibinfo{volume}{60}}, \bibinfo{pages}{1633} (\bibinfo{year}{1999}).

\bibitem[{\citenamefont{Bon\ifmmode~\check{c}\else \v{c}\fi{}a et~al.}(2000)\citenamefont{Bon\ifmmode~\check{c}\else \v{c}\fi{}a, Katras\ifmmode~\breve{}\else \u{}\fi{}nik, and Trugman}}]{bonca2000prl}
\bibinfo{author}{\bibfnamefont{J.}~\bibnamefont{Bon\ifmmode~\check{c}\else \v{c}\fi{}a}}, \bibinfo{author}{\bibfnamefont{T.}~\bibnamefont{Katras\ifmmode~\breve{}\else \u{}\fi{}nik}}, \bibnamefont{and} \bibinfo{author}{\bibfnamefont{S.~A.} \bibnamefont{Trugman}}, \bibinfo{journal}{Phys. Rev. Lett.} \textbf{\bibinfo{volume}{84}}, \bibinfo{pages}{3153} (\bibinfo{year}{2000}), \urlprefix\url{https://link.aps.org/doi/10.1103/PhysRevLett.84.3153}.

\bibitem[{\citenamefont{Ku et~al.}(2002)\citenamefont{Ku, Trugman, and Bon\ifmmode~\check{c}\else \v{c}\fi{}a}}]{bonca2002}
\bibinfo{author}{\bibfnamefont{L.-C.} \bibnamefont{Ku}}, \bibinfo{author}{\bibfnamefont{S.~A.} \bibnamefont{Trugman}}, \bibnamefont{and} \bibinfo{author}{\bibfnamefont{J.}~\bibnamefont{Bon\ifmmode~\check{c}\else \v{c}\fi{}a}}, \bibinfo{journal}{Phys. Rev. B} \textbf{\bibinfo{volume}{65}}, \bibinfo{pages}{174306} (\bibinfo{year}{2002}), \urlprefix\url{https://link.aps.org/doi/10.1103/PhysRevB.65.174306}.

\bibitem[{\citenamefont{Bon\ifmmode~\check{c}\else \v{c}\fi{}a and Trugman}(2021)}]{bonca2021}
\bibinfo{author}{\bibfnamefont{J.}~\bibnamefont{Bon\ifmmode~\check{c}\else \v{c}\fi{}a}} \bibnamefont{and} \bibinfo{author}{\bibfnamefont{S.~A.} \bibnamefont{Trugman}}, \bibinfo{journal}{Phys. Rev. B} \textbf{\bibinfo{volume}{103}}, \bibinfo{pages}{054304} (\bibinfo{year}{2021}), \urlprefix\url{https://link.aps.org/doi/10.1103/PhysRevB.103.054304}.

\bibitem[{\citenamefont{Kova\ifmmode~\check{c}\else \v{c}\fi{} and Bon\ifmmode~\check{c}\else \v{c}\fi{}a}(2024)}]{Klemen2024}
\bibinfo{author}{\bibfnamefont{K.}~\bibnamefont{Kova\ifmmode~\check{c}\else \v{c}\fi{}}} \bibnamefont{and} \bibinfo{author}{\bibfnamefont{J.}~\bibnamefont{Bon\ifmmode~\check{c}\else \v{c}\fi{}a}}, \bibinfo{journal}{Phys. Rev. B} \textbf{\bibinfo{volume}{109}}, \bibinfo{pages}{064304} (\bibinfo{year}{2024}), \urlprefix\url{https://link.aps.org/doi/10.1103/PhysRevB.109.064304}.

\bibitem{SM} See Supplemental Material at [URL] for additional 2eARPES results and analysis.

\bibitem[{\citenamefont{Kova\ifmmode~\check{c}\else \v{c}\fi{} and Bon\ifmmode~\check{c}\else \v{c}\fi{}a}(2025)}]{KlemenPRB2025}
\bibinfo{author}{\bibfnamefont{K.}~\bibnamefont{Kova\ifmmode~\check{c}\else \v{c}\fi{}}} \bibnamefont{and} \bibinfo{author}{\bibfnamefont{J.}~\bibnamefont{Bon\ifmmode~\check{c}\else \v{c}\fi{}a}}, \bibinfo{journal}{Phys. Rev. B} \textbf{\bibinfo{volume}{111}}, \bibinfo{pages}{115133} (\bibinfo{year}{2025}), \urlprefix\url{https://link.aps.org/doi/10.1103/PhysRevB.111.115133}.

\bibitem[{\citenamefont{Krsnik et~al.}(2020)\citenamefont{Krsnik, Strocov, Nagaosa, Bari\ifmmode \check{s}\else \v{s}\fi{}i\ifmmode~\acute{c}\else \'{c}\fi{}, Rukelj, Yakubenya, and Mishchenko}}]{Krsnik2020}
\bibinfo{author}{\bibfnamefont{J.}~\bibnamefont{Krsnik}}, \bibinfo{author}{\bibfnamefont{V.~N.} \bibnamefont{Strocov}}, \bibinfo{author}{\bibfnamefont{N.}~\bibnamefont{Nagaosa}}, \bibinfo{author}{\bibfnamefont{O.~S.} \bibnamefont{Bari\ifmmode \check{s}\else \v{s}\fi{}i\ifmmode~\acute{c}\else \'{c}\fi{}}}, \bibinfo{author}{\bibfnamefont{Z.}~\bibnamefont{Rukelj}}, \bibinfo{author}{\bibfnamefont{S.~M.} \bibnamefont{Yakubenya}}, \bibnamefont{and} \bibinfo{author}{\bibfnamefont{A.~S.} \bibnamefont{Mishchenko}}, \bibinfo{journal}{Phys. Rev. B} \textbf{\bibinfo{volume}{102}}, \bibinfo{pages}{121108} (\bibinfo{year}{2020}), \urlprefix\url{https://link.aps.org/doi/10.1103/PhysRevB.102.121108}.

\bibitem[{\citenamefont{Burovski et~al.}(2008)\citenamefont{Burovski, Fehske, and Mishchenko}}]{Burovski2008}
\bibinfo{author}{\bibfnamefont{E.}~\bibnamefont{Burovski}}, \bibinfo{author}{\bibfnamefont{H.}~\bibnamefont{Fehske}}, \bibnamefont{and} \bibinfo{author}{\bibfnamefont{A.~S.} \bibnamefont{Mishchenko}}, \bibinfo{journal}{Phys. Rev. Lett.} \textbf{\bibinfo{volume}{101}}, \bibinfo{pages}{116403} (\bibinfo{year}{2008}), \urlprefix\url{https://link.aps.org/doi/10.1103/PhysRevLett.101.116403}.

\bibitem[{\citenamefont{Mishchenko et~al.}(2014)\citenamefont{Mishchenko, Nagaosa, and Prokof'ev}}]{Mishchenko2014}
\bibinfo{author}{\bibfnamefont{A.~S.} \bibnamefont{Mishchenko}}, \bibinfo{author}{\bibfnamefont{N.}~\bibnamefont{Nagaosa}}, \bibnamefont{and} \bibinfo{author}{\bibfnamefont{N.}~\bibnamefont{Prokof'ev}}, \bibinfo{journal}{Phys. Rev. Lett.} \textbf{\bibinfo{volume}{113}}, \bibinfo{pages}{166402} (\bibinfo{year}{2014}), \urlprefix\url{https://link.aps.org/doi/10.1103/PhysRevLett.113.166402}.

\bibitem[{\citenamefont{Tupitsyn et~al.}(2016)\citenamefont{Tupitsyn, Mishchenko, Nagaosa, and Prokof'ev}}]{Tupitsyn2016}
\bibinfo{author}{\bibfnamefont{I.~S.} \bibnamefont{Tupitsyn}}, \bibinfo{author}{\bibfnamefont{A.~S.} \bibnamefont{Mishchenko}}, \bibinfo{author}{\bibfnamefont{N.}~\bibnamefont{Nagaosa}}, \bibnamefont{and} \bibinfo{author}{\bibfnamefont{N.}~\bibnamefont{Prokof'ev}}, \bibinfo{journal}{Phys. Rev. B} \textbf{\bibinfo{volume}{94}}, \bibinfo{pages}{155145} (\bibinfo{year}{2016}), \urlprefix\url{https://link.aps.org/doi/10.1103/PhysRevB.94.155145}.

\bibitem[{\citenamefont{Mishchenko et~al.}(2018)\citenamefont{Mishchenko, De~Filippis, Cataudella, Nagaosa, and Fehske}}]{Mishchenko2018}
\bibinfo{author}{\bibfnamefont{A.~S.} \bibnamefont{Mishchenko}}, \bibinfo{author}{\bibfnamefont{G.}~\bibnamefont{De~Filippis}}, \bibinfo{author}{\bibfnamefont{V.}~\bibnamefont{Cataudella}}, \bibinfo{author}{\bibfnamefont{N.}~\bibnamefont{Nagaosa}}, \bibnamefont{and} \bibinfo{author}{\bibfnamefont{H.}~\bibnamefont{Fehske}}, \bibinfo{journal}{Phys. Rev. B} \textbf{\bibinfo{volume}{97}}, \bibinfo{pages}{045141} (\bibinfo{year}{2018}), \urlprefix\url{https://link.aps.org/doi/10.1103/PhysRevB.97.045141}.

\end{thebibliography}

\end{document}